\documentclass{llncs}

\usepackage{etex}
\usepackage{epsfig}
\usepackage{url}
\usepackage[normalem]{ulem}
\usepackage{schemepgm}
\usepackage{graphpap}
\usepackage{amsmath}
\usepackage{amssymb}
\usepackage{color,pstcol}
\usepackage{colordvi}
\usepackage{pstricks}
\usepackage{pst-text}
\usepackage{pst-node}
\usepackage{pst-rel-points}
\usepackage{pst-coil}
\usepackage{algorithmic}
\usepackage{algorithm}

\newcommand{\cmt}[1]{}

\newcommand{\bw}[1]{\ensuremath{\overline{#1}}} 
\newcommand{\cat}[2]{\ensuremath{#1#2}} 
\newcommand{\expshare}{\ensuremath{\mathcal{E\!E\!S}}} 

\newcommand{\loc}[1]{\ensuremath{\mbox{\sf Loc}[{#1}]}}
\newcommand{\cell}[1]{\ensuremath{\mbox{\sf Cell}[{#1}]}}

\newcommand{\acar}{\ensuremath{\mathbf{0}}}
\newcommand{\acdr}{\ensuremath{\mathbf{1}}}
\newcommand{\bcar}{\ensuremath{\bar\acar}}
\newcommand{\bcdr}{\ensuremath{\bar\acdr}}
\newcommand{\acarset}{\ensuremath{\lbrace\acar\rbrace}}
\newcommand{\acdrset}{\ensuremath{\lbrace\acdr\rbrace}}
\newcommand{\bcarset}{\ensuremath{\lbrace\bcar\rbrace}}
\newcommand{\bcdrset}{\ensuremath{\lbrace\bcdr\rbrace}}
\newcommand{\epsilonset}{\ensuremath{\lbrace\epsilon\rbrace}}

\newcommand{\Af}[2]{\ensuremath{\Afonly_{\!\!#1}^{\,#2}}}
\newcommand{\Afonly}{\ensuremath{\mathcal{S\!F}}}
\newcommand{\Ap}[2]{\ensuremath{\Aponly_{\!\!#1}^{\,#2}}}
\newcommand{\Aponly}{\ensuremath{\mathcal{S\!P}}}
\newcommand{\Ae}[3]{\ensuremath{\Aeonly(#1, #2, #3)}}
\newcommand{\Aeonly}{\ensuremath{\mathcal{S\!E}}}
\newcommand{\Aa}[2]{\ensuremath{\Aaonly(#1, #2)}}
\newcommand{\Aaonly}{\ensuremath{\mathcal{S\!S}}}
\newcommand{\Av}{\ensuremath{\mathcal{S}}}
\newcommand{\Aphi}{\ensuremath{\Av^\emptyset}}

\cmt{{

}}

\newcommand{\Uf}[2]{\ensuremath{\Ufonly_{\!#1}^{\,#2}}}
\newcommand{\Ufonly}{\ensuremath{\mathcal{I}}}
\newcommand{\Df}[2]{\ensuremath{\Dfonly_{\!#1}^{\,#2}}}
\newcommand{\Dfonly}{\ensuremath{\mathcal{D}}}

\newcommand{\Lf}[2]{\ensuremath{\Lfonly_{\!\!#1}^{\,#2}}}
\newcommand{\Lfonly}{\ensuremath{\mathcal{L\!F}}}
\newcommand{\Lp}[2]{\ensuremath{\Lponly_{\!\!#1}^{\,#2}}}
\newcommand{\Lponly}{\ensuremath{\mathcal{L\!P}}}
\newcommand{\Le}[3]{\ensuremath{\Leonly(#1, #2, #3)}}
\newcommand{\Leonly}{\ensuremath{\mathcal{L\!E}}}
\newcommand{\Lv}{\ensuremath{\mathcal{L}}}
\newcommand{\Lvphi}{\ensuremath{\Lv^\emptyset}}

\newcommand{\AVf}[2]{\ensuremath{\AVfonly_{\!\!#1}^{\,#2}}}
\newcommand{\AVfonly}{\ensuremath{\mathcal{A\!F}}}
\newcommand{\AVp}[2]{\ensuremath{\AVponly_{\!\!#1}^{\,#2}}}
\newcommand{\AVponly}{\ensuremath{\mathcal{A\!P}}}
\newcommand{\AVgp}[2]{\ensuremath{\AVgponly_{\!\!#1}^{\,#2}}}
\newcommand{\AVgponly}{\ensuremath{{\mathcal{A\!B\!P}}}}

\newcommand{\AVe}[3]{\ensuremath{\AVeonly(#1, #2, #3)}}
\newcommand{\AVeonly}{\ensuremath{\mathcal{A\!E}}}
\newcommand{\AVv}{\ensuremath{\mathcal{A}}}

\newcommand{\plus}{\cup}

\newcommand{\rightk}[1]{\ensuremath{\stackrel{\scriptstyle #1}{\rightarrow}}}
\newcommand{\rightstar}{\rightk{\star}}
\newcommand{\eqdef}{\ensuremath{=}}

\newcommand{\append}{\mbox{\sf append}}

\newcommand{\lista}{\mbox{\sf lst1}}
\newcommand{\listb}{\mbox{\sf lst2}}

\newcommand{\nfa}{\ensuremath{\mathbf{N}}}
\newcommand{\nfabar}{\ensuremath{\overline\nfa}}

\newcommand{\prim}{\ensuremath{P}}
\newcommand{\exit}{{\sf pgm}}
\newcommand{\gram}{\ensuremath{G}}

\newcommand{\TwoCells}[2]{%
\psset{unit=.25mm}
\begin{pspicture}(0,-2)(36,18)
\psframe(0,-5)(36,15)
\psline(18,-4)(18,15)
\putnode{z}{origin}{9}{5}{\rnode{#1}{}}
\putnode{z}{origin}{27}{5}{\rnode{#2}{}}
\end{pspicture}%
}

\newcommand{\MIF}{\mbox{\sf\bf IF}}
\newcommand{\MLET}{\mbox{\sf\bf LET}}
\newcommand{\MIN}{\mbox{\sf\bf IN}}

\newcommand{\candidates}[1]{\ensuremath{\mbox{\sf\bf Candidates}(#1)}}
\newcommand{\visible}[1]{\ensuremath{{\sf\bf SVars}(#1)}}

\newcommand{\updateenv}[3]{{\sf update(#1, #2, #3)}}
\newcommand{\pair}[2]{(#1, #2)}

\newcommand{\pia}{\ensuremath{\pi_{a}}}
\newcommand{\pib}{\ensuremath{\pi_{b}}}

\begin{document}
\title{Heap Reference Analysis for Functional Programs}
\author{Amey Karkare\thanks{Supported by Infosys Technologies Limited,
    Bangalore, under Infosys Fellowship Award.} \and Amitabha Sanyal
  \and Uday Khedker}
\date{}
\institute{Department of CSE, IIT Bombay\\ Mumbai, India
\\\email{\{karkare,as,uday\}@cse.iitb.ac.in}}
\maketitle
\begin{abstract}
  Current  garbage  collectors  leave  a lot  of  garbage  uncollected
  because  they conservatively  approximate  liveness by  reachability
  from program  variables.  In this  paper, we describe a  sequence of
  static  analyses  that  takes  as  input  a  program  written  in  a
  first-order,  eager functional  programming language,  and  finds at
  each program point the references to objects that are guaranteed not
  to  be used  in the  future.   Such references  are made  null by  a
  transformation pass.   If this makes the object  unreachable, it can
  be collected by the garbage  collector.  This causes more garbage to
  be  collected, resulting  in fewer  collections.   Additionally, for
  those garbage collectors which  scavenge live objects, it makes each
  collection faster.

  The interesting aspects of our method are both in the identification
  of the analyses  required to solve the problem and  the way they are
  carried  out.   We  identify   three  different  analyses  ---  {\em
    liveness}, {\em sharing} and {\em accessibility}.  In liveness and
  sharing   analyses,   the    function   definitions   are   analyzed
  independently of the  calling context.  This is achieved  by using a
  variable  to represent  the unknown  context of  the  function being
  analyzed  and setting up  constraints expressing  the effect  of the
  function  with  respect  to  the  variable.   The  solution  of  the
  constraints is a  summary of the function that  is parameterized with
  respect to a calling context  and is used to analyze function calls.
  As a  result we  achieve context sensitivity  at call  sites without
  analyzing the function multiple number of times.
\end{abstract}


\section{Introduction}
\label{sec:intro}

An object is dead at an execution instant if it is not used in future.
Ideally, garbage  collectors should reclaim all objects  that are dead
at the  time of  garbage collection.  However,  even state of  the art
garbage  collectors  are not  able  to  distinguish between  reachable
objects that are  live and reachable objects that  are dead. Therefore
they  conservatively approximate  the  liveness of  an  object by  its
reachability from a set of  locations called the {\em root set} (stack
locations   and  registers  containing   program  variables).    As  a
consequence, many  dead objects are  left uncollected.  This  has been
confirmed   by  empirical   studies   for  Haskell~\cite{rojemo96lag},
Scheme~\cite{karkare06effectiveness}                                and
Java~\cite{shaham00gc,shaham01heap,shaham02estimating}.

In this paper,  we consider a first order  functional language without
imperative features  and propose a  method to release dead  objects so
that they can be collected by  the garbage collector.  This is done by
detecting unused references  to objects and setting them  to null.  If
all references to the object  are nullified, then the dead objects may
become  unreachable  and may  be  claimed  by  garbage collector.   We
propose  three  analyses  to   obtain  the  information  required  for
nullification: {\em liveness} analysis, which computes live references
at each program point (i.e.  the references used by the program beyond
the program  point), {\em sharing} analysis,  which computes alternate
ways to access live  references and {\em accessibility} analysis which
ensures that the references used by the nullification statement itself
exist  and  do  not  cause  a  dereferencing  exception.   An  earlier
paper~\cite{karkare07liveness} outlined the  basic method and provided
details of  the liveness analysis. This  paper brings the
theoretical aspects of the method to completion.

As our analyses  are interprocedural in scope, the  effect of function
calls on the heap must be modeled precisely. Most program analyses are
either not scalable  because they analyze the same  function more than
once or imprecise because they make overly safe worst-case assumptions
about the  effect of  a function  on the heap.   For a  better balance
between scalability and precision, one can compute context independent
summaries of  the effect of  functions on the  heap and then  use this
summary     at      particular     calling     context      of     the
function~\cite{chatterjee99relevant,whaley99compositional,%
cherem07fast}.  We do this by using a variable to represent an unknown
context  of the  function being  analyzed and  setting  up constraints
expressing the  effect of the  function with respect to  the variable.
The set of constraints is viewed as  a set of CFGs and the solution of
these constraints is a set  of finite state machines approximating the
languages defined  by the CFGs.  The  solution, which is  a summary of
the function parameterized with respect  to a calling context, is used
to analyze function calls.

The main contributions  of the paper are as  follows.  We identify the
analysis required  to find nullable references at  each program point.
As part of the analyses,  we show how context independent summaries of
functions  can be  obtained by  setting up  a set  of  constraints and
solving them by viewing them as  a CFG. Finally we show how the result
can  be  used for  safe  insertion  of  nullifying statements  in  the
program.

\vspace{-2mm}
\subsection{Motivation}
\label{sec:motiv}

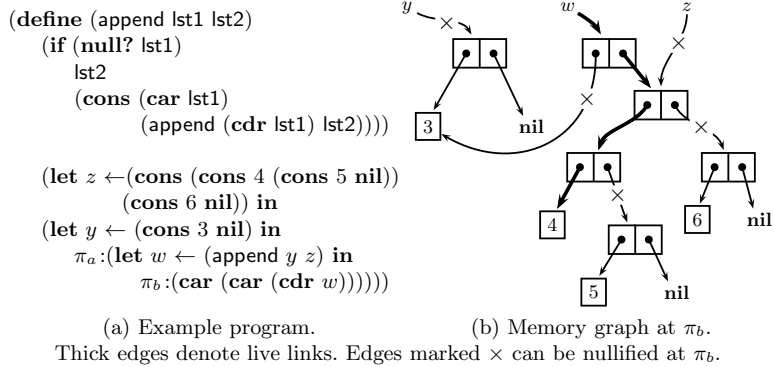
\begin{figure}[t]
  \hfill\scalebox{.90}{
      \begin{tabular}{@{}c@{}c@{}}
	\psset{unit=1mm}
	\begin{pspicture}(0,0)(60,45)
	  \rput(30,23){
	    \begin{uprogram}
	      \UFL\ \hspace*{-1\TAL} (\DEFINE\ (\append\  \lista\ \listb)
	      \UNL{0}  (\SIF\ (\NULLQ\ \lista)
	      \UNL{1}      \listb
	      \UNL{1}      (\CONS\ (\CAR\  \lista)
	      \UNL{3}           (\append\ (\CDR\  \lista) \listb))))
	      \UNL{0} 
	      \UNL{0} (\LET\ $z$\  $\leftarrow$(\CONS\ (\CONS\ $4$ (\CONS\ $5$ \NIL))
	      \UNL{2}\ \ \  (\CONS\ $6$ \NIL)) \IN
	      \UNL{0} (\LET\ $y$\  $\leftarrow$ (\CONS\ $3$ \NIL) \IN
	      \UNL{1}\  $\pia\!:$(\LET\ $w$\ $\leftarrow$\ (\append\ $y$\ $z$)\ \IN
	      \UNL{3}               $\pib\!:$(\CAR\ (\CAR\ (\CDR\
	      $w$))))))
	  \end{uprogram}}
	\end{pspicture}
	&
      \scalebox{.90}{\psset{unit=.85mm}
      \psset{linewidth=.3mm}
      \begin{pspicture}(0,0)(69,55)
	\putnode{o}{origin}{13}{50}{\TwoCells{o1}{o2}}
	\putnode{a}{o}{-10}{-15}{\psframebox{3}}
	\putnode{b}{o}{10}{-15}{\psframebox[linestyle=none,framesep=.5]{\NIL}}
	\ncline[offsetB=-.5,nodesepB=.1]{*->}{o1}{a}
	\ncline[offsetB=-.5,nodesepB=.1]{*->}{o2}{b}
	\putnode{y}{o}{-14}{8}{\psframebox[linestyle=none,framesep=.5]{$y$}}
	\nccurve[nodesepB=-.2,angleA=330,angleB=120]{->}{y}{o}
	\aput[-3.5](.5){\scalebox{1.2}{\psframebox[framesep=.2,linestyle=none,
	      fillstyle=solid,fillcolor=white]{$\times$}}}
	\putnode{c}{o}{25}{0}{\TwoCells{c1}{c2}}
	\putnode{d}{c}{10}{-10}{\TwoCells{d1}{d2}}
	\putnode{e}{d}{-13}{-12}{\TwoCells{e1}{e2}}
	\putnode{f}{d}{13}{-12}{\TwoCells{f1}{f2}}
	\ncline[nodesepB=-.5]{*->}{c2}{d}
	\ncline[nodesepB=-.5,linewidth=.7]{->}{c2}{d}
	\nccurve[ncurv=1,angleA=270,angleB=330]{*->}{c1}{a}
	\aput[-3.5](.2){\scalebox{1.2}{\psframebox[framesep=.2,linestyle=none,
	      fillstyle=solid,fillcolor=white]{$\times$}}}
	\nccurve[nodesepB=-.5,angleA=240,angleB=70]{*->}{d1}{e}
	\nccurve[nodesepB=-.5,angleA=240,angleB=70,linewidth=.7]{->}{d1}{e}
	\nccurve[nodesepB=-.5,angleA=300,angleB=110]{*->}{d2}{f}
	\aput[-3.5](.5){\scalebox{1.2}{\psframebox[framesep=.2,linestyle=none,
	      fillstyle=solid,fillcolor=white]{$\times$}}}
	\putnode{w}{c}{-8}{8}{\psframebox[linestyle=none,framesep=.2]{$w$}}
	\putnode{ww}{c}{15}{8}{\psframebox[linestyle=none,framesep=.2]{$z$}}
	\nccurve[nodesepB=-.2,angleA=330,angleB=120,linewidth=.7]{->}{w}{c}
	\putnode{g}{e}{-8}{-12}{\psframebox{4}}
	\putnode{h}{e}{8}{-14}{\TwoCells{h1}{h2}}
	\putnode{i}{f}{-6}{-11}{\psframebox{6}}
	\putnode{j}{f}{6}{-11}{\psframebox[linestyle=none,framesep=.5]{\NIL}}
	\ncline[offsetB=-.5,nodesepB=.1]{*->}{e1}{g}
	\ncline[offsetB=-.5,nodesepB=.1,linewidth=.7]{->}{e1}{g}
	\ncline[offsetB=-.5,nodesepB=-.3]{*->}{e2}{h}
	\aput[-3.2](.5){\scalebox{1.2}{\psframebox[framesep=.1,linestyle=none,
	      fillstyle=solid,fillcolor=white]{$\times$}}}
	\ncline[offsetB=-.5,nodesepB=.1]{*->}{f1}{i}
	\ncline[offsetB=-.5,nodesepB=.1]{*->}{f2}{j}
	\nccurve[nodesepB=-.2,angleA=270,angleB=90]{->}{ww}{d}
	\aput[-3.2](.4){\scalebox{1.2}{\psframebox[framesep=.1,linestyle=none,
	      fillstyle=solid,fillcolor=white]{$\times$}}}
	\putnode{k}{h}{-8}{-11}{\psframebox{5}}
	\putnode{l}{h}{8}{-11}{\psframebox[linestyle=none,framesep=.5]{\NIL}}
	\ncline[offsetB=-.5,nodesepB=.1]{*->}{h1}{k}
	\ncline[offsetB=-.5,nodesepB=.1]{*->}{h2}{l}
      \end{pspicture}} \\
      (a) Example program.&
      (b) Memory graph at $\pib$. \\
      \multicolumn{2}{c}{
	Thick edges denote live links.
	Edges marked $\times$ can be nullified at $\pib$.}\\
    \end{tabular}}
\hfill\mbox{}
\vskip -3mm
\caption{Example Program and its Memory Graph.}\label{fig:mot-exmp}    
\end{figure}

Figure~\ref{fig:mot-exmp}(a)  shows an  example  program.  The  label
$\pi$ of an expression $e$  denotes the program point just before the
evaluation of $e$.
The heap  memory can  be viewed as  a (possibly  unconnected) directed
acyclic  graph called {\em  memory graph}\footnote{Since  the language
under consideration (Sec.~\ref{sec:defs}) does not have any imperative
features, the memory graph can  not have cycles.}  during any instant
in the  execution of  the program.  The elements of  root set  are the
entry points for the memory graph.   The nodes in the memory graph are
the \CONS\ cells allocated in the heap.  There are three kind of edges
in the memory  graph: (1) Entry edges from an element  of the root set
to a  heap node,  (2) edges  from the \CAR\  field of  a heap  node to
another, and (3) edges from the \CDR\ field of a heap node to another.
Elements of the basic data  types and the 0-ary constructor \NIL\ form
the leaf nodes  of the graph.  All data is assumed  to be boxed, i.e.\
stored in  heap cells and  accessed through references.  The  edges in
the graph  are also called  {\em links}.  Figure~\ref{fig:mot-exmp}(b)
shows the memory graph at $\pib$.

The edges shown  by thick arrows are those  which will be dereferenced
beyond $\pib$.  These  edges are {\em live} at  $\pib$. Edges that are
not  live can  be  nullified  by the  compiler  by inserting  suitable
statements.  These edges are shown  with a $\times$ in the figure.  If
an object becomes  unreachable due to nullification of  such edges, it
can be collected by the garbage collector.  Note that an edge need not
be nullified if nullifying some  other edges makes it unreachable from
the root  set.  

To  find  out all  nullable  edges  in a  memory  graph,  we need  the
following analyses:
\begin{itemize}
\item For every  program point $\pi$, liveness analysis  finds out all
  the edges in  the memory graph that can be  potentially dereferenced
  along  some   path  from  $\pi$   to  exit.   For  the   program  in
  Fig.~\ref{fig:mot-exmp}, the edges  corresponding to references $w$,
  $(\CDR\; w)$, $(\CAR\; (\CDR\;  w))$, $(\CAR\; (\CAR\; (\CDR\; w)))$
  should be marked as live at $\pib$.
\item Sharing analysis is used to identify all possible ways to access
  live edges.  In Fig.~\ref{fig:mot-exmp}, the expression $(\CAR\; z)$
  is not directly used beyond $\pib$.  However, sharing analysis gives
  us that $z$ and $(\CDR\; w)$  share a \CONS\ cell. Therefore, we can
  not nullify $(\CAR\; z)$, as the edge is live due to use of $(\CAR\;
  (\CDR\; w)$.  Using sharing analysis, we infer that the complete set
  of  expressions  corresponding to  live  edges  at  $\pib$ is:  $w$,
  $(\CDR\; w)$, $(\CAR\; (\CDR\; w))$, $(\CAR\; (\CAR\; (\CDR\; w)))$,
  $(\CAR\; z)$, $(\CAR\; (\CAR\; z))$.
\item Since  our analysis is static,  it is possible that  not all the
  \CONS\ cells in the sequence of links that we dereference to nullify
  a non-live link  have been created during a  particular execution of
  the program.   This can happen if  a \CONS\ cell in  the sequence of
  links is allocated in one branch of a conditional expression and not
  in  the other.   Accessibility analysis  ensures that  the statement
  used for nullification  does not dereference a \CONS\  cell which is
  not allocated.
\end{itemize}

\vspace{-1mm}
\subsection{Organization}
\label{sec:organization}

Section~\ref{sec:defs} describes the language  used  to explain  our
analysis along with the basic concepts and notations.
Liveness analysis is described in Section~\ref{sec:deref}.
Section~\ref{sec:sharing} explains the analysis to compute sharing
between root variables.
Section~\ref{sec:availability} describes availability analysis.
%
Section~\ref{sec:null-ins}  describes the actual process of null
insertion.
The related work is given in Section~\ref{sec:rel-work}. We conclude in
Section~\ref{sec:concl} and give the direction for future research.

\section{Concepts and Notations}
\label{sec:defs}
The  syntax of  our  language is  shown  in Fig.~\ref{fig:lang}.   The
language  has call-by-value  semantics. The  argument  expressions are
evaluated from left to right.  We assume that variables in the program
are renamed so  that the same name is not  defined in multiple scopes.
The body of the program is the expression denoted by $e_\exit$.
We write $\pi\!:\!e$ to associate $\pi$ with
the program point just before the expression $e$.
\begin{figure}[t]
    \hfill\scalebox{.95}{$\begin{array}{lcr}
      p & ::= & d_1 \ldots d_n \; e_\exit \;\;\;\;\;\;\;\;\;
      \;\;\;\;\;\;\;\;\;\;\;\;\;\;
      \mbox{\em --- program}\\
      d & ::= & (\DEFINE\; (f\; v_1 \; \ldots \;v_n)\; e_1) \;\;\;\;\;\; 
      \mbox{\em --- function definition} \\ 
      e & ::= & \;\;\;\;\;\;\;\;\;\;\;\;\;\;\;\;\;\;\;\;\;\;\;\;\;\;
      \;\;\;\;\;\;\;\;\;\;\;\;\;\;
      \mbox{\em --- expression}\\
      & & \begin{array}{lll}
        \;\; \kappa && \mbox{\em --- constant }\\
        \mid v && \mbox{\em --- variable} \\
	\mid \NIL & \mid (\CONS\; e_1\; e_2) & \mbox{\em ---  constructors} \\
	\mid (\CAR\; e_1) & \mid (\CDR\; e_1) & \mbox{\em --- selectors} \\
	\mid (\PAIRQ\; e_1) & \mid (\NULLQ\; e_1) & \mbox{\em ---  testers} \\
	\mid (\PRIM\; e_1\; e_2) && \mbox{\em --- generic primitive} \\
	\mid (\SIF\; e_1\; e_2\; e_3) && \mbox{\em --- conditional} \\
	\mid (\LET\; v_1 \leftarrow e_2\; \IN\; e_3)
	&& \mbox{\em --- let binding} \\
	\mid (f\; e_1\;\ldots\; e_n) && \mbox{\em --- function application}
      \end{array}
     \end{array}$}\hfill\mbox{}
\vskip -2mm %
  \caption{The syntax of our language}\label{fig:lang}
\end{figure}

An edge emanating  from a \CAR\ field is labeled  \acar\ while an edge
emanating from  a \CDR\  field is labeled  \acdr.  Entry edges  do not
have any label.  There are  two kinds of traversals associated with an
edge: A {\em forward} traversal is in the direction of the edge, and a
{\em backward}  traversal is  in the opposite  direction of  the edge.
For  an  edge  with  label  $l (l\in  \{\acar,  \acdr\})$,  a  forward
traversal  over the edge  is denoted  by $l$,  while \bw{l}  denotes a
backward traversal over the edge.

Given a node in  a memory graph, a {\em path} is  a sequence of labels
representing a  traversal over connected  edges starting at  the node.
In  general,  a  path  involves  both  forward  as  well  as  backward
traversals  over edges.  A  {\em forward}  path involves  only forward
traversals  over  edges,  and  a  {\em backward}  path  involves  only
backward traversals over edges. Note  that starting from a \CONS\ cell
there  can be  multiple  possible edge  traversals  labeled \bcar\  or
\bcdr, but at most one traversal labeled \acar\ or \acdr.  In general,
all forward traversals  from a node have unique  labels while multiple
backward traversals may share the  same label. A {\em bipath} consists
of  a (possibly  empty) forward  path followed  by a  (possibly empty)
backward path.  Note that forward and backward paths are special cases
of bipath.  Only bipaths are important to us because liveness, sharing
and accessibility  can all be  described using bipaths.  We  use Greek
letters   ($\alpha$,   $\beta$,   \ldots)   to  denote   paths.    The
concatenation of two path segments  $\alpha$ and $\beta$ is denoted by
\cat{\alpha}{\beta}.   The   reverse  of  a   path  $\alpha$,  denoted
\bw{\alpha}, is the path which  traverses the edges of $\alpha$ in the
opposite order and  opposite direction.  We have: \mbox{$\bw{\epsilon}
  =    \epsilon$},    and    \mbox{$\bw{\cat{\alpha_1}{\alpha_2}}    =
  \cat{\bw{\alpha_2}}{\     \bw{\alpha_1}}$}.     The    concatenation
($\sigma_1\cdot\sigma_2$) of a set of paths $\sigma_1$ with $\sigma_2$
is  defined as  a  set  containing concatenation  of  each element  in
$\sigma_1$ with each element in $\sigma_2$.

A path can be simplified by repeatedly removing consecutive occurrences
of  backward and  forward  traversal  of the  same  edge (in  general,
removing occurrences of \cat{\bw{\alpha}}{\alpha}).  The reduction does
not change the  semantics of the path in that the  node reached by the
path remains  the same even  after simplification.  Further,  since we
are  interested  in  bipaths  only, paths  containing  \bcdr\acar\  or
\bcar\acdr\  can be  ignored.  This  gives  us the  following rules  of
reduction:
\begin{eqnarray}
  \scalebox{.95}{$\begin{array}{lcr@{\hspace{1.5cm}}lcr@{\hspace{1.5cm}}lcr}
  \alpha_1\bcar\acar\alpha_2 & \rightarrow & \alpha_1\alpha_2 &
  \alpha_1\bcdr\acdr\alpha_2 & \rightarrow & \alpha_1\alpha_2 &
  \alpha_1\bot\alpha_2 & \rightarrow & \bot \\
  \alpha_1\bcar\acdr\alpha_2 & \rightarrow & \bot &
  \alpha_1\bcdr\acar\alpha_2 & \rightarrow & \bot &
  \end{array} \label{eqn:eup-red}$}
\end{eqnarray}

$\alpha  \rightk{k}  \alpha'$ denotes  the  reduction  of $\alpha$  to
$\alpha'$  in $k$ steps,  and $\rightstar$  denotes the  reflexive and
transitive closure of $\rightarrow$.  A  path which can not be reduced
further using above rules is said  to be in {\em canonical} form. Note
that a path in canonical form is either a bipath or $\bot$.

Very often we shall be interested in paths that start from a heap cell
pointed directly by  the root set.  We call such  paths as {\em access
  paths}.  Let \loc{e} denote the stack location which holds the value
of $e$\footnote{For a  root variable $r$, \loc{r} is  same as $r$. For
  any other expression $e$, \loc{e} can be thought of as the temporary
  that holds the  value of $e$.}  and $\cell{e}$  denote the heap node
pointed to by $\loc{e}$. We use $e.\alpha$ to denote access path which
starts in the  heap at $\cell{e}$ and traverse  the path $\alpha$.  If
$\sigma$ denotes a set of paths,  then $e.\sigma$ is the set of access
paths rooted at $\cell{e}$  corresponding to $\sigma$. i.e.  $e.\sigma
\eqdef \lbrace e.\alpha \mid \alpha \in \sigma \rbrace $ We use access
paths to refer  to links in the memory graph. The  link referred to by
an access path is the last edge in a traversal using the access path.

The syntax of the meta-language  used to describe our analysis is very
similar to  the language being analyzed. To  distinguish between them,
the keywords in the meta-language  are written in all capitals (\MLET,
\MIN, \MIF\ etc.).


\section{Liveness Analysis}
\label{sec:deref}

A link  in a memory  graph is  live at a  program point $\pi$  if some
expression dereferences  it beyond $\pi$.   An access path is  live if
the link denoted by it is live.   Note that an access path can be live
in two ways: either it is used directly to access the last link, or it
shares the live link with some  other access path using which the link
is made  live. Liveness analysis discovers access  paths through which
the live link is directly accessed.

The  {\em   liveness  environment\/}  at   $\pi$,  denoted  $\Lv_\pi$,
describes all the  live access paths at $\pi$.  It  is a function from
root variables to  sets of paths.  The result  of liveness analysis is
the annotation of each program point with its liveness environment.
The liveness of  an access path before an  expression $e$ depends upon
its use inside  $e$ itself and in the rest of  the program through the
result  of  $e$.  Therefore  we  define  a  transfer function  denoted
\Leonly\ to compute the liveness of access paths before an expression,
given the liveness of result after the expression.  As expressions may
contain applications  of primitive  operations and functions,  we also
need to  propagate liveness across  these applications.  This  is done
through  the  summarizing   functions  \Lponly\  and  \Lfonly.   While
\Lponly\ is  given directly based  on the semantics of  the primitive,
\Lfonly\ is inferred from the body of a function.

\vspace{-1mm}
\subsection{Liveness Transfer Function (\Leonly)}
\label{sec:compute_ue}
For an expression $\pi\!:\!e$, a  set of paths $\sigma$ specifying the
liveness of the result of  evaluating $e$ and the liveness environment
\Lv\ after $e$,  \Le{e}{\sigma}{\Lv} computes the liveness environment
at $\pi$.   The liveness environment  associated with the exit  of any
function is  empty liveness  environment $\Lvphi$ defined  as $\forall
x\; \Lvphi(x)  = \emptyset$.  The liveness associated  with the result
of the program expression  $e_\exit$ is $\sigma_\exit = \lbrace \acar,
\acdr \rbrace  *$, i.e.\ the entire  result of the  program is needed.
For any other function $f$, the liveness associated with the result is:

\scalebox{.95}{$\begin{array}{rcl}
  \sigma_{exit_f}   &=&   \!\!\!\!\!\!\!
  \displaystyle{\bigcup_{\mbox{\scriptsize all calls }(f\; e_1\;\ldots\;e_n)}}
  \!\!\!\!\!\!\!
  \left\{ \sigma \mid \mbox {$\sigma$ is
    the  liveness  of the result of  $(f\;  e_1\;  \ldots\;  e_n)$ after
    the call}
  \right\}
\end{array}$}

\begin{figure}[t]
\begin{eqnarray}
    \Le{\kappa}{\sigma}{\Lv}  &=& \Lv \label{eqn:lv_kappa}\\
    \Le{v}{\sigma}{\Lv}  &=& \updateenv{\Lv}{v}{\Lv(v)\cup\sigma}
    \label{eqn:lv_var}\\
    \Le{(\SIF\; e_1\; e_2\; e_3)}{\sigma}{\Lv} 
    &=& \nonumber(\MLET\; \Lv' \leftarrow \Le{e_3}{\sigma}{\Lv}\;\MIN \\
    & & \nonumber\;(\MLET\; \Lv'' \leftarrow \Le{e_2}{\sigma}{\Lv}\;\MIN \\
    & & \;\; \Le{e_1}{\epsilonset}{\Lv' \cup \Lv''}))
    \label{eqn:lv_if}  \\ 
    \Le{(\LET\; v_1 \leftarrow e_1\; \IN\; e_2)}{\sigma}{\Lv}
    &=& \nonumber(\MLET\; \Lv' \leftarrow \Le{e_2}{\sigma}{\Lv}\;\MIN \\
    & & \;\; \Le{e_1}{\Lv'(v_1)}{\updateenv{\Lv'}{v_1}{\emptyset}})
    \label{eqn:lv_let} \\
    \begin{array}[t]{r}
      \Le{(\prim\; {e_1}\;\ldots\; {e_n})}{\sigma}{\Lv}\\
      \mbox{ }\prim\mbox{ is a primitive} 
    \end{array}
    &=& \nonumber\begin{array}[t]{l}
      (\MLET\;\Lv_1\leftarrow\Le{e_n}{\Lp{\prim}{n}(\sigma)}{\Lv}\;\MIN \\
      \;(\MLET\;\Lv_2\leftarrow\Le{e_{n-1}}{\Lp{\prim}{n-1}(\sigma)}{\Lv_1}\;
      \MIN \\
      \;\;\;\;\ldots  \end{array}\\
     & & \;\;\;\;\;(\MLET\;\Lv_{n-1}\leftarrow\Le{e_2}{\Lp{\prim}{2}(\sigma)}{\Lv_{n-2}}\;
      \MIN \nonumber \\
     & & \;\;\;\;\;\;\;\;\Le{e_1}{\Lp{\prim}{1}(\sigma)}{\Lv_{n-1}}))\ldots)
      \label{eqn:lv_prim}\\
    \begin{array}[t]{r}
      \Le{(f\; e_1\ldots e_n)}{\sigma}{\Lv}\\
      \mbox{ }f\mbox{ is a user defined function} 
    \end{array}
    &=& \nonumber\begin{array}[t]{l}
      (\MLET\;\Lv_1\leftarrow\Le{e_n}{\Lf{f}{n}(\sigma)}{\Lv}\;\MIN \\
      \;(\MLET\;\Lv_2\leftarrow\Le{e_{n-1}}{\Lf{f}{n-1}(\sigma)}{\Lv_1}\;\MIN \\
      \;\;\;\;\ldots  \end{array} \\
    & & \;\;\;\;\;(\MLET\;\Lv_{n-1}\leftarrow\Le{e_2}{\Lf{f}{2}(\sigma)}{\Lv_{n-2}}\;
    \MIN \nonumber \\
    & & \;\;\;\;\;\;\;\;\Le{e_1}{\Lf{f}{1}(\sigma)}{\Lv_{n-1}}))\ldots)
    \label{eqn:lv_fun}
  \end{eqnarray}
\vskip -5mm %
\caption{Computing \Leonly}\label{fig:le}
\end{figure}

The  computation of \Leonly\  is given  in Fig.~\ref{fig:le}.\footnote{
{\sf update} is a helper function to compute change in environments:

$\begin{array}{rcl}
  {\sf \updateenv{OldEnv}{Y}{NewVal}(X)} &=& {\sf (\MIF\ (X == Y)\ NewVal\ 
    (OldEnv\ X))}
\end{array}$
} In  the expression \mbox{$(\SIF\;  e_1\; e_2\; e_3)$},  the $\sigma$
for $e_1$ is \epsilonset\ because the value of $e_1$ is used to decide
the branch, for which  only \cell{e_1} is used (\ref{eqn:lv_if}).  For
a \LET, the liveness of $v_1$  from $e_2$ and beyond is transferred to
$e_1$ (\ref{eqn:lv_let}).  The liveness environment before a primitive
application  is computed by  using \Lponly\  to transfer  the liveness
from  the  result  of  the   application  to  each  of  its  arguments
(\ref{eqn:lv_prim}).  Similarly, applications of user defined functions
use \Lfonly (\ref{eqn:lv_fun}).

As the result of liveness  analysis is the annotation of every program
point   with   its  liveness   environment,   during  computation   of
\Le{e}{\sigma}{\Lv}, the  program point  before $e$ is  annotated with
the computed  liveness environment  as a side  effect. We do  not show
this explicitly to avoid clutter.
\subsection{Summarizing Functions (\Lponly\ and \Lfonly)}
\label{sec:compute_up_uf}

If $\sigma$ describes the set  of paths specifying the liveness of the
result   of  $(\prim\   e_1  \ldots   e_n)$  after   the   call,  then
\Lp{\prim}{i}($\sigma$) gives  the set of access  paths specifying the
liveness of $e_i$ at the program point after $e_i$.
The summarizing functions for the primitives in our language, \CAR, \CDR,
\CONS,  \NULLQ,  \PAIRQ\  and  \PRIM,  are  shown  below.   The  0-ary
constructor \NIL\ does not accept any argument and is ignored.
\begin{eqnarray}
\scalebox{.95}{$
\begin{array}{rcl@{\hspace{2cm}}rcl}
  \Lp{\CAR}{1}(\sigma)   &=& \epsilonset \plus \acarset\cdot\sigma &
  \Lp{\CDR}{1}(\sigma)   &=& \epsilonset \plus \acdrset\cdot\sigma \\
  \Lp{\CONS}{1}(\sigma)  &=& \bcarset\cdot\sigma &
  \Lp{\CONS}{2}(\sigma)  &=& \bcdrset\cdot\sigma \\
  \multicolumn{6}{c}{\Lp{\NULLQ}{1}(\sigma) = \epsilonset, \hspace{5mm}
  \Lp{\PAIRQ}{1}(\sigma) = \epsilonset,  \hspace{5mm}
  \Lp{\PRIM}{1}(\sigma)  = \epsilonset, \hspace{5mm}
  \Lp{\PRIM}{2}(\sigma)  = \epsilonset  }
\end{array}$}
\end{eqnarray}

$\Lp{\CAR}{1}(\sigma)$ includes $\acarset\cdot\sigma$ because the link
described by a path labeled $\alpha$ from $\cell{(\CAR\; e)}$ can also
be described by the  path labeled $\acar\alpha$ from $\cell{e}$. Also,
as the cell corresponding to $e$ is used to find the value of \CAR, we
need to add $\epsilon$ to  the live paths of $e$.  Reasoning
about (\CDR\ $e$) is similar.
For similar reasons, a path $\alpha$ describing the liveness of \CONS\
translates   to  an   $\bcar\alpha$  for   its  first   argument,  and
$\bcdr\alpha$ for  its second argument.   Further, as \CONS\  does not
read its arguments,  the access paths of the  arguments do not contain
$\epsilon$.
The  remaining  primitives  read  only  the value  of  the  arguments,
therefore the set of live path of the arguments is $\epsilonset$.

\Lfonly\ plays the  same role as \Lponly\ for  user defined functions.
Given  a  function  defined  as \mbox{($\DEFINE\;  (f\;  v_1\;\ldots\;
v_n)\; e) $} and a $\sigma$ specifying the set of paths specifying the
liveness of the result, \Lfonly\ is computed as follows:
\begin{equation}
  \Lf{f}{i}(\sigma) =  \Le{e}{\sigma}{\emptyset}(v_i), \;  1\leq i\leq n
  \label{eqn:lv_lfonly}
\end{equation}

\begin{example}\label{exmp:motivation_analysis}
  To  compute   the  transfer   functions  for  \append,   we  compute
  \Le{e}{\sigma}{\emptyset} in terms of a variable $\sigma$.  Here $e$
  is the body of \append.  Figure~\ref{fig:xe_append} shows the values
  at various program points in \append.  From the liveness information
  of the parameters \lista\ and \listb, we get: \newcommand{\lbeleven}{\mbox{\listb.\Lf{\append}{2}($\bcdrset\cdot\sigma$)}}
\newcommand{\leleven}{$\lbrace \lbeleven \rbrace$}
\newcommand{\laten}{\mbox{$\lista.\acdrset\cdot\Lf{\append}{1}(\bcdrset\cdot\sigma)$}}
\newcommand{\lten}{$\left\lbrace\begin{array}{c}\laten,\\ \lbeleven\end{array}\right\rbrace$}
\newcommand{\laseven}{\mbox{$\lista.(\epsilonset \cup \lbrace\acar\bcar\rbrace\cdot\sigma \cup \acdrset\cdot\Lf{\append}{1}(\bcdrset\cdot\sigma))$}}
\newcommand{\lseven}{$\left\lbrace\begin{array}{c}\laseven,\\ \lbeleven\end{array}\right\rbrace$}
\newcommand{\lbfour}{\mbox{$\listb.(\sigma \cup \Lf{\append}{2}(\bcdrset\cdot\sigma))$}}
\newcommand{\lfour}{$\left\lbrace\begin{array}{c}\laseven,\\ \lbfour\end{array}\right\rbrace$}
\newcommand{\ap}[1]{
  \psframebox[linestyle=none,fillcolor=lightgray,fillstyle=solid,framesep=.5]{%
  \bf #1}}

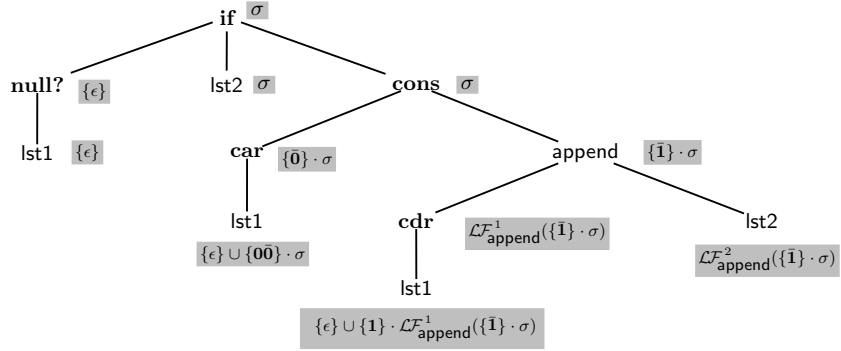
\begin{figure}[t]
\hfill\scalebox{.90}{
\psset{unit=1mm}
\begin{pspicture}(0,2)(\textwidth,54)
  \putnode{if}{origin}{32}{50}{\SIF}
  {\putnode{apif}{if}{4}{1}{\ap{$\sigma$}}}
  \putnode{cond}{if}{-28}{-10}{\NULLQ}
  {\putnode{apcond}{cond}{8}{-1}{\scalebox{.8}{\ap{\epsilonset}}}}
  
  \putnode{true}{if}{0}{-10}{\listb}
  {\putnode{aptrue}{true}{5}{0}{\ap{$\sigma$}}}
  
  \putnode{false}{if}{28}{-10}{\CONS}
  {\putnode{apfalse}{false}{7}{0}{\ap{$\sigma$}}}
  
  \ncline[nodesep=1]{-}{if}{cond}
  \ncline[nodesep=1]{-}{if}{false}
  \ncline[nodesep=1]{-}{if}{true}
  
  \putnode{argc}{cond}{0}{-10}{\lista}
  {\putnode{apargc}{argc}{7}{0}{\scalebox{.8}{\ap{\epsilonset}}}}
    \ncline[nodesep=.1]{cond}{argc}
  
  \putnode{acons}{false}{-25}{-10}{\CAR}
  {\putnode{apacons}{acons}{8}{-1}{\scalebox{.8}{
	\ap{$\bcarset\cdot\sigma$}}}}
  \putnode{bcons}{false}{25}{-10}{\append}
  {\putnode{apbcons}{bcons}{12}{0}{\scalebox{.8}{
	\ap{$\bcdrset\cdot\sigma$}}}}
  \ncline[nodesep=.2]{false}{acons}
  \ncline[nodesep=.2]{false}{bcons}
  
  \putnode{acar}{acons}{0}{-10}{\lista}
  {\putnode{apacar}{acar}{0}{-5}{\scalebox{.8}{
	\ap{$\epsilonset\cup\{\acar\bcar\}\cdot\sigma$}}}}
  \ncline[nodesep=.2]{acons}{acar}
  
  \putnode{a0app}{bcons}{-25}{-10}{\CDR}
  {\putnode{apa0app}{a0app}{17}{-2}{\scalebox{.8}{
	\ap{$\Lf{\append}{1}(\bcdrset\cdot\sigma)$}}}}
  \putnode{bapp}{bcons}{26}{-10}{\listb}
  {\putnode{apbapp}{bapp}{0}{-6}{\scalebox{.8}{
	\ap{$\Lf{\append}{2}(\bcdrset\cdot\sigma)$}}}}
  \ncline[nodesep=.2]{a0app}{bcons}
  \ncline[nodesep=.2]{bapp}{bcons}
  
  \putnode{bcdr}{a0app}{0}{-10}{\lista}
  {\putnode{apbcdr}{bcdr}{0}{-6}{\scalebox{.8}{
	\ap{\ap{$\epsilonset\cup\acdrset\cdot\Lf{\append}{1}(\bcdrset\cdot\sigma)$}}}}}
  \ncline[nodesep=.2]{bcdr}{a0app}
\end{pspicture}}
\hfill\mbox{}
\caption{Transformation of access paths for body of \append\label{fig:xe_append}}
\end{figure}

\scalebox{.95}{$\begin{array}[b]{rcl}
  \Lf{\append}{1}(\sigma)
  &=& \epsilonset \plus \lbrace\acar\bcar\rbrace\cdot\sigma\;
  \plus\; \acdrset\cdot\Lf{\append}{1}(\bcdrset\cdot\sigma)\\ 
  \Lf{\append}{2}(\sigma)
  &=& \sigma \plus \Lf{\append}{2}(\bcdrset\cdot\sigma) \\
\end{array}$}
\qed\end{example}

\subsection{Solving Liveness Equations}
We now  describe briefly  the steps to  solve the  liveness equations.
The         reference~\cite{karkare07liveness}         and         the
Appendix~\ref{sec:solving_eqns}   both  contain  a   detailed  example
illustrating these steps.  Further, the equations resulting out of the
sharing analysis are also solved in a similar manner.

In  general, the  equations defining  the   functions \Lfonly\
will be recursive.  To solve  such equations we start by guessing that
the solution  for $\Lf{f}{i}(\sigma)$ will be of  the form: $\Uf{f}{i}
\plus \Df{f}{i}\cdot\sigma$, where \Uf{f}{i} and \Df{f}{i} are sets of
strings over the alphabet $\lbrace \acar, \acdr,\bcar, \bcdr \rbrace$.
Then,
\begin{enumerate}
\item We substitute the guessed form of \Lf{f}{i} in the equations and
 equate the  $\sigma$-dependent and $\sigma$-independent  parts of LHS
 and RHS of  each equation. This gives us  equations for \Uf{f}{i} and
 \Df{f}{i} which are independent of $\sigma$.

\item We  interpret the equations as  rules of a  context free grammar
 (CFG)  with \Uf{f}{i}  and  \Df{f}{i} as  non-terminals.  The set  of
 terminal  symbols   of  the  CFG  is   $\lbrace\acar,  \acdr,  \bcar,
 \bcdr\rbrace$.

\item We add more rules to represent the liveness at different program
  points in terms of the above non-terminals.

\item  We approximate the  CFG by  a set  of non  deterministic finite
  automata (NFA) and simplify the  NFAs so that the paths in canonical
  form are accepted.  The algorithm describing this step and its proof
  of  correctness   is  given  in   Appendix~\ref{sec:nfa-elim}.   The
  algorithm  is  a  revised  version  of that  given  in  our  earlier
  work~\cite{karkare07liveness}.
\end{enumerate}


\section{Sharing Analysis}
\label{sec:sharing}

Given  a memory  graph, expressions  $e_1$ and  $e_2$ are  involved in
sharing if there are  forward paths from $\cell{e_1}$ and $\cell{e_2}$
to a common heap cell.  In particular we are interested in the sharing
of the root variables. Let $h$ be a heap cell shared by root variables
$x$ and $y$.  Let the forward access path $x.\alpha$ describe the path
from $x$  to $h$  and the forward  access path $y.\beta$  describe the
path from $y$  to $h$.  Then, sharing between $x$ and  $y$ can be seen
as a bipath labeled  $\alpha\bw{\beta}$ from \cell{x}\ to \cell{y}\ in
the  memory graph.   Fig.~\ref{fig:sharing-examp} shows  some  ways in
which sharing can arise.

\begin{figure}[t]
  \hfill
  \scalebox{.92}{\begin{tabular}{|l|c|c|c|}\hline
      \rule{0pt}{15pt}{\bf Expression:} &
      \begin{tabular}{l} 
	$(\LET\; y_1 \leftarrow (\CAR\; x_1)\; \IN\;$\\
	$\;\;\;\;\;\pi_1\!:\ldots)$ 
      \end{tabular}
      &
      \begin{tabular}{l} 
	$(\DEFINE\; (f\; v_1\; v_2)\;\; \pi_2:\ldots)$\\
	\ldots\\
	$(f\; (\CAR\; x_2)\; x_2)$
      \end{tabular}&
      $(\LET\; y_3 \leftarrow (\CONS\; x_3\; x_3)\; \IN\; \pi_3\!:\ldots)$\\\hline
      \rule{0pt}{15pt}{\bf Sharing:} & 
      $\acar \in \Av_{\pi_1}(x_1, y_1)$&
      $\bcar \in \Av_{\pi_2}(v_1, v_2)$&
      $\acar\bcdr, \acdr\bcar \in \Av_{\pi_3}(y_3, y_3);$
      $\bcar, \bcdr \in \Av_{\pi_3}(x_3, y_3) $ \\ \hline
    \end{tabular}}
  \hfill\mbox{}
\vskip -3mm %
\caption{Examples of sharing\label{fig:sharing-examp}}
\end{figure}

The {\em  sharing environment} at $\pi$,  denoted $\Av_\pi$, describes
the sharing between root variables  in any memory graph that can arise
at $\pi$.  The sharing environment is  a function from pairs of root
variables to  sets of bipaths.  The result of  sharing analysis is to
annotate  each program  point  with an  approximation  of its  sharing
environment.

Since  variables  take their  values  from  evaluation of  expressions
(through \LET\  or argument  bindings), it is  convenient to  define a
function denoted \Aeonly, which computes the sharing between variables
and  expressions.    Further,  we  also  need   to  propagate  sharing
environments  across  applications of  primitive  operations and  user
defined functions.   This is done  by using the  summarizing functions
\Aponly\ and  \Afonly.  For  a primitive \prim,  \Ap{\prim}{i} denotes
the sharing  between the  $i^{th}$ argument and  the result  of \prim.
\Af{f}{i}  is  interpreted  in  a similar  manner.   Additionally  the
function \Aaonly\ computes the sharing of an expression with itself.

\subsection{Sharing Transfer Function (\Aeonly)}
\label{sec:compute_ae}
The transfer function \Ae{x}{e}{\Av} computes the extent  of sharing between
the  root   variable  $x$  and  the  result   obtained  by  evaluating
$e$.\footnote{The  function  can  easily  be  extended  to  a  set  of
variables so that only a single pass over the expression is required.}
In  our language,  the  sharing  between root  variables  can only  be
affected either  at the  \LET-binding or at  the entry of  a function.
The  computation  of \Aeonly\  begins  at  function definitions.   The
sharing  environment before  the program  expression $e_\exit$  is the
empty sharing  environment $\Aphi$ defined as  $\forall x,y\; \Aphi(x,
x) = \epsilonset,  \Aphi( x, y) = \emptyset$.   For any other function
$f$, defined  as $(\DEFINE\;  (f\; v_1\ldots\;v_n)\; e)$,  the initial
sharing    environment     $\Av_{entry_f}$    is    as     shown    in
Fig.~\ref{fig:sharing-entry}.
\begin{figure}[t]
\begin{eqnarray*}
  \Av_{entry_f}(v_i, v_i) &=& \epsilonset \cup\;
  \displaystyle{\bigcup_{\pi:(f\; e_1\; \ldots\; e_n)}}
  \Aa{e_i}{\Av_\pi} \\
  \Av_{entry_f}(v_i, v_j) &=& 
  \displaystyle{\bigcup_{\pi:(f\; e_1\; \ldots\; e_n)}}
  \expshare(e_i, e_j, \Av_\pi, \visible{\pi}) \\
  \nonumber \mbox{where }
  && 1\leq i, j\leq n,\;\; i\not=j, \;\;
  \visible{\pi} = \mbox{set of root variables in scope at
  $\pi$}\nonumber \\
\expshare(e, e', \Av, \mbox{Vars}) &=& 
  \left\{\bw{\alpha}\beta \mid \alpha \in \Ae{x}{e}{\Av},
  \beta \in \Ae{x}{e'}{\Av}, x \in \mbox{Vars}
  \right\}
\end{eqnarray*}
\vskip -5mm 
\caption{Sharing at the entry of a function}\label{fig:sharing-entry}
\end{figure}

\begin{figure}[t]
\begin{eqnarray}
  \Ae{x}{\kappa}{\Av} &=& \emptyset \label{eqn:sharing_kvar}
  \\
  \Ae{x}{v}{\Av}  &=& \Av(x,v)
      \label{eqn:sharing_varvar}
  \\
  \Ae{x}{(\SIF\; e_1\; e_2\; e_3)}{\Av} &=&
  (\MLET\; \Av' \leftarrow \Ae{x}{e_1}{\Av}\; \MIN\; 
  \mbox{\hskip 1cm\{$\Av'$ is ignored\}}\nonumber\\
  && \;\;\;\; \Ae{x}{e_2}{\Av} \cup \Ae{x}{e_3}{\Av})
  \label{eqn:sharing_if}  \\
  \Ae{x}{(\LET\; v_1 \leftarrow e_1\; \IN\; e_2)}{\Av}
  &=& (\MLET\; \Av' \leftarrow \updateenv{\Av}{(v_1, v_1)}{\epsilonset\cup\Aa{e_1}{\Av}}
  \;\MIN \nonumber \\
  & & \;\;\;\;(\MLET\; \Av'' \leftarrow \updateenv{\Av'}{(x, v_1)}{\Ae{x}{e_1}{\Av}}\; \MIN \nonumber\\
  & & \;\;\;\;\;\;\;\; \Ae{x}{e_2}{\Av''}))   \label{eqn:sharing_let}
  \\
  \begin{array}[t]{r}
    \Ae{x}{(\prim\; {e_1}\;\ldots\; {e_n})}{\Av} \\
    \prim\mbox{ is a primitive}
  \end{array}
  &=& \displaystyle\bigcup_{1\leq i\leq n}
        \Ae{x}{e_i}{\Av}\cdot\Ap{\prim}{i}
	\label{eqn:sharing_prim} 
  \\
  \begin{array}[t]{r}
    \Ae{x}{(f\; e_1\;\ldots\; e_n)}{\Av} \\
    f\mbox{ is a user defined function}
  \end{array}
  &=& \displaystyle\bigcup_{1\leq i\leq n}
  \Ae{x}{e_i}{\Av}\cdot\Af{f}{i} \label{eqn:sharing_fun} 
\end{eqnarray}
\vskip -5mm %
\caption{Computing \Aeonly}\label{fig:ae}
\end{figure}

The  computation  of  \Ae{x}{e}{\Av}  is given  in  Fig.~\ref{fig:ae}.
Equations~(\ref{eqn:sharing_kvar})  and (\ref{eqn:sharing_varvar}) are
self-explanatory.   In an  \SIF\  expression, sharing  can  be due  to
execution  of either  branch.  The  sharing between  $x$ and  $e_1$ is
computed to  propagate the sharing  environment inside $e_1$;  it does
not affect  the sharing between $x$  and the \SIF\  expression.  For a
\LET\  expression, sharing  environment $\Av'$  at $e_2$  captures the
sharing between  $x$ and $v_1$  (\ref{eqn:sharing_let}).  Finally, the
sharing  between $x$  and the  result  of application  of a  primitive
\prim\ is obtained by composing the sharing between $x$ and $e_i$ with
the sharing between the $e_i$ and the result (\ref{eqn:sharing_prim}).
User defined functions  (\ref{eqn:sharing_fun}) are treated similarly.
Note  that only  \LET\  expression modifies  the sharing  environment.
During computation  of \Ae{x}{e}{\Av} the program point  before $e$ is
annotated  with \Av.   However, as  in \Leonly,  we do  not  show this
explicitly.

\subsection{Summarizing Functions (\Aponly\ and \Afonly)}
\label{sec:compute_af_ap}
\Aponly\ specifies the extent  of sharing between the formal arguments
of  a primitive  and its  return value.  The sharing  between $i^{th}$
argument and the  result is denoted by \Ap{\prim}{i}.  For a primitive
application $(\prim\ e_1 \ldots e_n)$:
\begin{eqnarray*}
  \alpha\bw{\beta} \in \Ap{\prim}{i} &\Rightarrow&
  \mbox{there is a bipath $\alpha\bw{\beta}$ from \loc{e_i} to 
  \loc{(\prim\ e_1 \ldots e_n)}}
\end{eqnarray*}
The  functions \Ap{\prim}{i},  of a  primitive are  computed  from its
semantics:
\begin{eqnarray}
\scalebox{.95}{$\begin{array}{r@{\ }c@{\ }l@{\hspace{.3cm}}r@{\ }c@{\
    }l@{\hspace{.3cm}}r@{\ }c@{\ }l@{\hspace{.3cm}}r@{\ }c@{\ }l}
  \Ap{\CAR}{1}     &=& \acarset &
  \Ap{\CDR}{1}     &=& \acdrset &
  \Ap{\CONS}{1}    &=& \bcarset &
  \Ap{\CONS}{2}    &=& \bcdrset \\
  \Ap{\NULLQ}{1} &=& \emptyset &
  \Ap{\PAIRQ}{1} &=& \emptyset &
  \Ap{\PRIM}{1} &=& \emptyset &
  \Ap{\PRIM}{2} &=& \emptyset
\end{array}$}
\end{eqnarray}


\Afonly\ specifies the extent  of sharing between the formal arguments
of  a function  and its  return value.   The sharing  between $i^{th}$
argument and  the result is  denoted by \Ap{\prim}{i}. For  a function
defined as  $ (\DEFINE\ (f\;  v_1\; \ldots\; v_n)\; e)$,  \Af{f}{i} is
computed as follows:
\begin{eqnarray}
  \Af{f}{i} &=& \Ae{v_i}{e}{\Aphi}, \;  1\leq i\leq n
  \label{eqn:alias-afonly} 
\end{eqnarray}

\subsection{Sharing with Self  (\Aaonly)}
\label{sec:sharing_self}
Because  of  the sharing  in  the  subexpressions,  the result  of  an
expression may share a \CONS\  cell along two different paths. We call
it self sharing,  and use the function \Aaonly\  to capture it. The
computation of \Aaonly\ is shown in Fig.~\ref{fig:aaonly}.
\begin{figure}[t]
\begin{eqnarray}
  \Aa{\kappa}{\Av} &=& \epsilonset \label{eqn:slf_kvar}
  \\
  \Aa{v}{\Av}  &=& \Av(v,v) \label{eqn:slf_varvar}
  \\
  \Aa{(\SIF\; e_1\; e_2\; e_3)}{\Av} &=&
  \Aa{e_2}{\Av} \cup \Aa{e_3}{\Av}  \label{eqn:slf_if}
  \\
  \begin{array}[t]{r}
    \Aa{(\LET\; v_1 \leftarrow \pi_1\!:\!e_1\; \IN\; e_2)}{\Av}\\
    \visible{\pi_1} = \{x_1, \ldots, x_n\}
  \end{array} 
  &=& \begin{array}[t]{l}
    (\MLET\; \Av_0 \leftarrow \updateenv{\Av}{(v_1, v_1)}{ \Aa{e_1}{\Av}}\; \MIN \nonumber\\
     \;(\MLET \Av_1 \leftarrow \updateenv{\Av_1}{(x_1, v_1)}{ \Ae{x_1}{e_1}{\Av}}\;
  \MIN\end{array} \nonumber\\
  && \;\;\;\;\ldots \nonumber\\
  && \;\;\;\;\;(\MLET \Av_n \leftarrow \updateenv{\Av_{n-1}}{(x_n, v_1)}{\Ae{x_n}{e_1}{\Av}}\;
  \MIN \nonumber\\
  & & \;\;\;\;\;\;\;\; \Aa{e_2}{\Av_n})\ldots))   \label{eqn:slf_let}
  \\
  \begin{array}[t]{r}
    \Aa{\pi\!:(\prim\; {e_1}\;\ldots\; {e_n})}{\Av} \\
    \prim\mbox{ is a primitive}
  \end{array}
  &=&\phantom{\cup}\displaystyle\bigcup_{\stackrel{1\leq i,j\leq n}{i\not=j}}
  \bw{\Ap{\prim}{i}}\cdot   (\displaystyle{\bigcup_{\pi}}
  \expshare(e_i, e_j, \Av, \visible{\pi}))
  \cdot\Ap{\prim}{j}\nonumber\\
  && {\cup}\displaystyle\bigcup_{1\leq i\leq n}
  \bw{\Ap{\prim}{i}}\cdot\Aa{e_i}{\Av}\cdot\Ap{\prim}{i} 
  \label{eqn:slf_prim} 
  \\
\begin{array}[t]{r}
    \Aa{\pi\!:(f\; e_1\;\ldots\; e_n)}{\Av} \\
    f\mbox{ is a user defined function}
  \end{array}
  &=&\phantom{\cup} \displaystyle\bigcup_{\stackrel{1\leq i,j\leq n}{i\not=j}}
  \bw{\Af{f}{i}}\cdot   (\displaystyle{\bigcup_{\pi}}
  \expshare(e_i, e_j, \Av,\visible{\pi}))
  \cdot\Af{f}{j} \nonumber \\
  && {\cup}\displaystyle\bigcup_{1\leq i\leq n}
  \bw{\Af{f}{i}}\cdot\Aa{e_i}{\Av}\cdot\Af{f}{i}
 \label{eqn:slf_fun}\end{eqnarray}
\vskip -5mm %
\caption{Computing \Aaonly}\label{fig:aaonly}
\end{figure}

\subsection{Computing Aliases of Access Paths}
\label{sec:aliasing}
We say that  two access paths are {\em aliased} at  a program point if
they  share the  same  \CONS\ cell  in  the heap  at  that point.   We
distinguish between  two kinds of  aliases: two access paths  are {\em
  link}-aliases if they share the last edge in the path, otherwise
they are {\em node}-aliases.

The result of sharing analysis can be used to compute all aliases of a
given access path at a given point.  Let $\pi$ be a program point, and
let  $\Av_\pi$  be the  sharing  environment  at  $\pi$. Further,  let
$x.\alpha$ be an access path  under consideration, where $\alpha$ is a
forward path.  To find out the aliases of $x.\alpha$ rooted at $y$, we
proceed as follows. Consider the set $\Av_\pi(y,x)$ which contains the
bipaths from $\cell{y}$ to  $\cell{x}$.  For $\beta \in \Av_\pi(y,x)$,
if $\beta\alpha$ reduces  to a forward path then  $y.\beta\alpha$ is a
forward access path which reaches the same \CONS\ cell as that reached
by $x.\alpha$ implying that $y.\beta\alpha$ is an alias of $x.\alpha$.
Because we do  not have the bipaths in  $\Av_pi$ explicitly listed, we
have to compute CFGs describing the bipaths. This is same as described
for               liveness               (App.~\ref{sec:solving_eqns},
\cite{karkare07liveness}). We also  compute the trivial CFG describing
the   string   $\alpha$.   The   concatenation   of   CFG   describing
$\Av_\pi(y,x)$ with CFG describing the string $\alpha$ gives a CFG,
which after conversion to NFA and simplification gives the regular
grammar describing  the aliases of $x.\alpha$ rooted at $y$.

The  link alias  of a  root variable  $x$ is  $x$ itself.  To  get the
link-aliases of  $x.\alpha\acar$, we compute aliases  of $x.\alpha$ as
described above, and extend it by \acar. Similarly we can compute
link-aliases for $x.\alpha\acdr$.


\section{Accessibility Analysis}
\label{sec:availability}

To nullify a link $l$ at a program point $\pi$, we have to traverse an
access path  from some root variable,  say $v$, to the  source cell of
$l$. However, it  is possible that some \CONS\ cell  $c$ in the access
path from $v$ to $l$ is  created along one execution path to $\pi$ but
not along another.   Since the nullification of $l$  at $\pi$ requires
the cell $c$ to be dereferenced, a run time exception may occur if the
execution path  taken is the one  along which $c$ is  not created.  To
avoid  this, we  need  to make  sure  that the  access  path used  for
nullification  is such  that  all  the intermediate  cells  in it  are
definitely created.

\begin{example}\label{exmp:mot-avail}
  Consider the following program fragment:

  \hfill  \begin{uprogram}  \UFL $(\LET\  x  \leftarrow  (\SIF\; (y  <
    5)\;\; (\CONS\; 2\;  z)\;\; \NIL)\; \IN\;\;\; \pi\!:\!\!(\SIF\; (y
    \ge 5)\;\;\;\; \pi_1\!: w\;\;\;\; \pi_2\!:\!\!(\CDR\; x))$
  \end{uprogram}\hfill\mbox{}

  Observe that  the program does not raise  a dereferencing exception.
  Assume that the link $x.0$ is not live at $\pi$. This information is
  not sufficient to  nullify $x.0$ safely at $\pi$  because it does not
  guarantee  that variable  $x$  points  to a  \CONS\  cell at  $\pi$.
  Similarly, knowing that $x.0$ is not live at $\pi_1$ or $\pi_2$ does
  not  enable us  to nullify  $x.0$ at  those points.   However, since
  $(\CDR\;  x)$  dereferences  $x$,  we  can infer  that  $x$  can  be
  dereferenced  at $\pi_2$.   Thus,  we can  safely  nullify $x.0$  at
  $\pi_2$.
\hfill\qed\end{example}

Assuming that  the program cannot generate  a dereferencing exception,
it  is  possible  to  infer  the  set of  access  paths  that  can  be
dereferenced  without  causing exception.   We  call  such paths  {\em
  accessible}.   There are  two ways  in which  the set  of accessible
paths can be inferred at $\pi$.  We can discover access paths in which
all  the \CONS\  cells are  either created  or dereferenced  along all
program paths from the program entry to $\pi$.  We call these paths as
{\em available} paths at $\pi$. Secondly, we can discover access paths
in which all the \CONS\ cells are dereferenced along all program paths
from  $\pi$  to  the  program  exit.   We call  these  paths  as  {\em
  anticipable}.

In this  paper we describe  availability analysis only. The  result of
availability analysis  is the {\em  availability environment}, denoted
\AVv, corresponding to each program  point. It is a function from root
variables  to  the  corresponding  available access  paths.   We  next
describe how the availability environment is computed.

\vspace{-3mm}
\subsection{Availability Transfer Function (\AVeonly)}
\label{sec:compute_ave}

In  general,   an  expression   has  to  dereference   the  structures
corresponding to  its subexpressions.  Therefore, for  execution to
proceed  normally,   these  structures  must  exist.    We  call  this
requirement as the  {\em demand} on a subexpression.  We  use a set of
paths  to  describe  the   demand.   The  demand  from  the  enclosing
expression   is  modified  by   an  expression   and  passed   to  its
subexpressions.

\begin{example}
  \CAR\ and \CDR\ require that their arguments are non null. Thus, for
  expression $(\CAR\;\; \pi_1\!:\!(\CDR\;\; \pi_2\!:\!x))$, the demand
  at $\pi_1$ is \epsilonset\ due  to the \CAR\ application. The demand
  at $\pi_2$  is $\{  \epsilon, \acdr\}$, where  $\epsilon$ is  due to
  \CDR, and  \acdr\ is because  of the demand of \CAR\ on $(\CDR\; x)$
  which is modified and passed to $x$.
\hfill\qed\end{example}  

The  way availability information  is generated  and propagated  is as
follows. Consider an expression $(\CAR\; (\CDR\; x))$. Assume that the
availability  environment  before this  expression  indicates that  no
access  path  rooted   at  $x$  is  available.   When   we  reach  the
subexpression  $x$, the  chain of  selectors $(\CAR\;  (\CDR\; \ldots$
generates the demand $\{\epsilon, \acdr\}$  on $x$. We thus update the
availability  environment of $x$  to include  $x.\{\epsilon, \acdr\}$.
This availability is propagated upwards  and used to conclude that the
availability of $(\CDR\; x)$ is $\epsilon$. Thus availability analysis
involves  a  inward  propagation  of  demand followed  by  an  outward
propagation of availability.

Given an expression  $e$, the set of access  paths $\sigma$ describing
the demand on the result of $e$ and the availability environment \AVv\
at the  program point before $e$,  we compute the  availability of $e$
and the availability environment after $e$ using the transfer function
\AVeonly.  This  is  described  in  Fig.~\ref{fig:ave}.   Availability
analysis  is  an all  paths  problem.   We  get constraints  involving
intersection   operation   for   sets  describing   availability.   As
intersection operation can not be  mapped directly to CFGs, we need to
get  an approximate  (but  safe)  solution.  This  is  achieved by  an
intraprocedural analysis in which we neither propagate the demand from
function application to its  arguments, nor propagate the availability
of  arguments  to  the  function application  (\ref{eqn:avv_fun}).   A
straightforward unfolding  of \AVeonly\ will give  us the availability
environment at different program points.
\begin{figure}[t]
  \begin{eqnarray}
    \AVe{\kappa}{\sigma}{\AVv}  &=& \pair{\epsilonset}{\AVv}
    \label{eqn:avv_kappa}\\
    \AVe{v}{\sigma}{\AVv} 
    &=& (\MLET\; \AVv' \leftarrow \updateenv{\AVv}{v}{\AVv(v) \cup \sigma}\;
    \MIN\nonumber \\ 
    && \;\; \pair{\AVv'(v)}{\AVv'})\label{eqn:avv_var}\\
    \AVe{(\SIF\; e_1\; e_2\; e_3)}{\sigma}{\AVv} 
    &=& \nonumber
    (\MLET\; \pair{\sigma_1}{\AVv_1} \leftarrow \AVe{e_1}{\epsilonset}{\AVv}\;\MIN \\
    && \nonumber\;\;
    (\MLET\; \pair{\sigma_2}{\AVv_2} \leftarrow \AVe{e_2}{\sigma}{\AVv_1}\;\MIN \\
    && \nonumber\;\;\;\;
    (\MLET\; \pair{\sigma_3}{\AVv_3} \leftarrow \AVe{e_3}{\sigma}{\AVv_1}\;\MIN \\
    & & \;\;\;\;\;\;\pair{\sigma \cup (\sigma_2 \cap \sigma_3)}{\AVv'}))) \label{eqn:avv_if}  \\
    \mbox{where } \AVv'(v)&=& \AVv_2(v) \cap \AVv_3(v)\;\nonumber    \\ 
    \AVe{(\LET\; v_1 \leftarrow e_1\; \IN\; e_2)}{\sigma}{\AVv}
    &=& \nonumber
    (\MLET\; \pair{\sigma'}{\AVv'} \leftarrow \AVe{e_1}{\emptyset}{\AVv}\;\MIN \\
    & &\; \AVe{e_2}{\sigma}{\updateenv{\AVv'}{v_1}{\sigma'}})
    \label{eqn:avv_let}  \\
    \begin{array}[t]{r}
      \AVe{(\prim\; {e_1}\;\ldots\; {e_n})}{\sigma}{\AVv}\\
      \mbox{ }\prim\mbox{ is a primitive} 
    \end{array}
    &=& \begin{array}[t]{l}
      (\MLET\;\pair{\sigma_1}{\AVv_1}\leftarrow\AVe{e_1}{\AVp{\prim}{1}(\sigma)}{\AVv}\;\MIN \\
      \;(\MLET\;\pair{\sigma_2}{\AVv_2}\leftarrow\AVe{e_{2}}{\AVp{\prim}{2}(\sigma)}{\AVv_1}\;
      \MIN \\
      \;\;\;\;\ldots \\
      \;\;\;\;\;(\MLET\;\pair{\sigma_{n}}{\AVv_{n}}\leftarrow
      \AVe{e_{n}}{\AVp{\prim}{{n}}(\sigma)}{\AVv_{n-1}}\; \MIN \\
      \;\;\;\;\;\;\pair{\sigma\; \cup \;\displaystyle\bigcup_{1\leq i\leq n}\!\!
	\AVgp{\prim}{i}(\sigma_i)}{\AVv_n})\ldots))
    \end{array} 
    \label{eqn:avv_prim}\\
    \begin{array}[t]{r}
      \AVe{(f\; e_1\ldots e_n)}{\sigma}{\AVv}\\
      \mbox{ }f\mbox{ is a user defined function} 
    \end{array}
    &=&
    \begin{array}[t]{l}
      (\MLET\;\pair{\sigma_1}{\AVv_1}\leftarrow\AVe{e_1}{\emptyset}{\AVv}\;\MIN
      \\
      \;(\MLET\;\pair{\sigma_2}{\AVv_2}\leftarrow\AVe{e_{2}}{\emptyset}{\AVv_1}\;\MIN
      \\
      \;\;\;\;\ldots \\
      \;\;\;\;\;(\MLET\;\pair{\sigma_n}{\AVv_{n}}\leftarrow
      \AVe{e_{n}}{\emptyset}{\AVv_{n-1}}\; \MIN \\
      \;\;\;\;\;\;\pair{\sigma}{\AVv_n})\ldots))
    \end{array} \label{eqn:avv_fun}
  \end{eqnarray}
\vskip -5mm %
\caption{Computing \AVeonly}\label{fig:ave}
\end{figure}

\vspace{-5mm}
\subsection{Inward Propagation of Demand  (\AVponly)}
\label{sec:avp_avf}

If $\sigma$  describes the set of  paths specifying the  demand on the
result  of evaluating  the primitive  application $(\prim\  e_1 \ldots
e_n)$   then   \AVp{\prim}{i}($\sigma$)  gives   the   set  of   paths
representing the  demand on $e_i$.   \cmt{{\AVponly\ is in  most cases
identical  to  \Lponly\ because  both  the  analyses  answer the  same
question:  Given the substructure  of the  result to  be dereferenced,
what   is  the   substructure   of  the   $i^{th}$   argument  to   be
dereferenced?}}
\begin{eqnarray}
\scalebox{.95}{$
\begin{array}{rcl@{\hspace{2cm}}rcl}
  \AVp{\CAR}{1}(\sigma)   &=& \epsilonset \plus \acarset\cdot\sigma &
  \AVp{\CDR}{1}(\sigma)   &=& \epsilonset \plus \acdrset\cdot\sigma \\
  \AVp{\CONS}{1}(\sigma)  &=& \bcarset\cdot\sigma &
  \AVp{\CONS}{2}(\sigma)  &=& \bcdrset\cdot\sigma \\
  \multicolumn{6}{c}{\AVp{\NULLQ}{1}(\sigma) = \emptyset, \hspace{5mm}
  \AVp{\PAIRQ}{1}(\sigma) = \emptyset,  \hspace{5mm}
  \AVp{\PRIM}{1}(\sigma)  = \emptyset, \hspace{5mm}
  \AVp{\PRIM}{2}(\sigma)  = \emptyset  }
\end{array}$}
\end{eqnarray}

 \cmt{{ The
function  \AVfonly\ propagates  demands through  function applications
and is interpreted in a similar manner. It is defined as:
\begin{eqnarray}
  \AVf{f}{i}(\sigma) &=&  (\MLET\; (\sigma', \AVv') \leftarrow
  \AVe{e}{\sigma}{\emptyset}\; 
  \MIN \; \AVv'(v_i)),
  \;  1\leq i\leq n,
  \label{eqn:av_afonly}
\end{eqnarray}
}}

\subsection{Outward Propagation of Availability (\AVgponly)}
\label{sec:avgp_f}

If  $\sigma$  describes  the  availability  of  $i^{th}$  argument  of
$(\prim\  e_1 \ldots e_n)$,  then $\AVgp{\prim}{i}(\sigma)$  gives the
availability of $(\prim\ e_1 \ldots  e_n)$.  For the primitives in our
language:
\begin{eqnarray}
\scalebox{.95}{$\begin{array}{rcl@{\hspace{2cm}}rcl}
    \AVgp{\CAR}{1}(\sigma)   &=& \bcarset\cdot\sigma &
    \AVgp{\CDR}{1}(\sigma)   &=& \bcdrset\cdot\sigma \\
    \AVgp{\CONS}{1}(\sigma)  &=& \epsilonset\cup\acarset\cdot\sigma &
    \AVgp{\CONS}{2}(\sigma)  &=& \epsilonset\cup\acdrset\cdot\sigma \\
    \multicolumn{6}{c}{
      \AVgp{\NULLQ}{1}(\sigma) = \emptyset, \hspace{5mm}
      \AVgp{\PAIRQ}{1}(\sigma) = \emptyset, \hspace{5mm}
      \AVgp{\PRIM}{1}(\sigma)  = \emptyset, \hspace{5mm}
      \AVgp{\PRIM}{2}(\sigma)  = \emptyset
    }
  \end{array}$}
\end{eqnarray}

\newcommand{\nullifiable}{\mbox{\sf Nullifiable}}
\newcommand{\properprefix}{\mbox{\sf ProperPrefix}}
\section{Null Insertion}
\label{sec:null-ins}
We need to consider the following issues for null insertion:
\begin{itemize}
\item {\em  Safety:} No live  edge should be nullified.   Further, the
  expression used  to nullify  an edge should  not dereference  a null
  reference.
\item {\em  Profitability:} An  edge should be  nullified as  early as
  possible.  Multiple nullification  of  same edge,  through the  same
  expression or through its aliases, should be avoided.
\end{itemize}


Safety, can be achieved by the following:
\begin{enumerate}
\item The  proper prefixes of  the access path used  for nullification
  should be  available.  Thus, the  candidate access paths at  a given
  program point  can be obtained  by extending available  access paths
  with a  \acar\ and  a \acdr.  Additionally,  all root  variables are
  also  candidates for  null insertion.   The liveness  of  only these
  paths need to be checked for null insertion. Thus,

  \hfill\scalebox{.95}{$\begin{array}{rcl}
    \candidates{\pi} &=& 
    \displaystyle\bigcup_{v \in \visible{\pi}} v.(\epsilonset \cup \{\alpha\acar,
    \alpha\acdr \mid \alpha \in \AVv_\pi(v)\}) 
  \end{array}$}\hfill\mbox{}

\item  To  make sure  that  the link  described  by  a candidate  path
  $v.\alpha$  is  not  live,  we  have  to  compute  link  aliases  of
  $v.\alpha$ and ensure that none of them is live at $\pi$.
\end{enumerate}
\vskip -2mm
  
Our analyses annotate  liveness, sharing and availability environments
at the  program points before the expressions.   Therefore, we nullify
dead links  at these program points  only. The analyses  can easily be
extended to compute  the environments at the program  points after the
expressions, so that links can be nullified at these points.

To address the profitability issue, we visit the program points in the
order of execution (reverse  depth-first order of the expression tree)
to nullify links.  We mark the access paths which are already used for
nullification, and do not nullify them again.  However, redundant null
insertions are still  possible because the same link  may be nullified
more than  once through  aliased access paths.   In general it  is not
possible  to eliminate  redundant  null insertions.   However, we  can
reduce them  by computing {\em  must-aliases} that hold on  all paths,
and  marking  all  must-link-aliases  of  the  access  path  used  for
nullification.

\label{sec:how-null}
\newcommand{\set}{\mbox{\sf\bf SET!}}
\newcommand{\setcar}{\mbox{\sf\bf SET-CAR!}}
\newcommand{\setcdr}{\mbox{\sf\bf SET-CDR!}}
\newcommand{\linknull}{\mbox{\sf LinkNullify}}
\newcommand{\nullify}{\mbox{\sf Nullify}}
\newcommand{\sbegin}{\mbox{\sf\bf BEGIN}}

A  given nullifiable  access path  can be  translated  into equivalent
expression for nullification of the  link it represents. We need three
primitives   in  our   meta  language   to  achieve   the   effect  of
nullification. These are: \set\  to nullify root variable, \setcar\ to
nullify \CAR\  references, and  \setcdr\ to nullify  \CDR\ references.
The expression  for nullification from  access path is  obtained using
the  function  \nullify\  which  is inserted  at  appropriate  program
points:

\hfill\scalebox{.95}{
$\begin{array}{rcl}
  \nullify(v.\alpha) &=& \left\{\begin{array}{l@{\hspace{5mm}}l}
      (\set\; v\; \NIL) & \alpha=\epsilon\\
      \linknull(\alpha, v) & \alpha\not=\epsilon\\
    \end{array}\right. \\
  \linknull(1\alpha, e) &=&
  \left\{\begin{array}{l@{\hspace{5mm}}l}
      (\setcdr\; e\; \NIL) & \alpha=\epsilon\\
      \linknull(\alpha, (\CDR\; e)) & \alpha\not=\epsilon
    \end{array}\right.\\
  \linknull(0\alpha, e) &=&
  \left\{\begin{array}{l@{\hspace{5mm}}l}
      (\setcar\; e\; \NIL) & \alpha=\epsilon\\
      \linknull(\alpha, (\CAR\; e))  & \alpha\not=\epsilon
    \end{array}\right.
\end{array}$
}\hfill\mbox{}
\vspace{-2mm}
\section{Related Work}
\label{sec:rel-work}

Existing   literature  regarding   improving  memory   usage   can  be
categorized as follows:

\noindent{\em     Compile    time     reuse}.     The     method    by
Barth~\cite{barth77shifting} detects memory  cells with zero reference
count and reallocates them for  further use in the program.  Jones and
Le~Metayer~\cite{jones89compile}  describe  a  sharing analysis  based
garbage collection for reusing of cells which collects a cell provided
expressions using it do not need it for their evaluation.

\noindent{\em      Explicit       reclamation}.       Shaham      et.\
al.~\cite{shaham05establishing}  use  an  automaton called  {\em  heap
  safety automaton\/} to model safety of inserting a free statement at
a   given   program  point.    The   analysis   is   based  on   shape
analysis~\cite{sagiv02shape} and  is very precise. However  it is very
inefficient.
{\em   Free-Me}~\cite{guyer06free}  combines  a   lightweight  pointer
analysis  with  liveness  information  that detects  when  short-lived
objects die and  insert statements to free such  objects. The analysis
is  simpler and  cheaper as  the  scope is  limited. 
The  analysis  described   by  Inoue  et.\  al.~\cite{inoue88analysis}
detects the scope (function) out  of which a cell becomes unreachable,
and explicitly claims the cell whenever the execution goes out of that
scope.   Like  our  method,  the  result of  their  analysis  is  also
represented  using CFGs. The  main difference  between their  work and
ours is  that we  detect and nullify  dead links  at any point  of the
program, while they detect and collect objects that are unreachable at
function boundaries.

\noindent{\em  Making  dead objects  unreachable}.   The most  popular
approach  to  make  dead  objects  unreachable  is  to  identify  live
variables    and    reduce    the    root   set    to    only    these
variables~\cite{agesen98garbage}.   The drawback  of this  approach is
that  all heap  objects reachable  from  the live  root variables  are
considered live, even if some of them remain unused.
{\em  Escape   analysis}  \cite{Blanchet:2003:EAJ,choi99escape}  based
approaches discover objects escaping  a procedure, i.e.\ objects whose
lifetimes outlive  the procedure that created  them.  All non-escaping
objects  are  allocated  on  stack, whereby  they  become  unreachable
whenever the creating procedure exits.
{\em  Region}  based  garbage collection~\cite{tofte02combining}  uses
{\em region inference\/}~\cite{tofte98region} to identify regions that
are allocated storage for objects.  Memory blocks are always allocated
in a particular region and are deallocated at the end of that region's
lifetime.  Escape analysis and  region inference
detect garbage only  at the boundaries of certain  predefined areas of
the program.
In  our  previous  work~\cite{khedker06heap},  we  have  used  bounded
abstractions of  access paths called  {\em access graphs}  to describe
the liveness of memory links in imperative programs and have used this
information to  nullify dead  links. This paper  is completion  of our
earlier  work~\cite{karkare07liveness},  where  we  used  liveness  to
introduce the ideas presented in this paper.


\vspace{-2mm}
\section{Conclusions and Future Work}
\label{sec:concl}
\vspace{-2mm}

In  this paper  we have  proposed a  method to  nullify links  in heap
memory to improve garbage collection.  The method consists of a set of
analyses to  discover dead references at every  program point followed
by the actual insertion of null statements. We claim that the analyses
are both  scalable and precise---scalable because we  obtain a context
dependent  summary of  each  function call,  and  precise because  the
summaries are  used in a context- and flow-sensitive analysis of
each function call. The method  is very similar to the {\em functional
  method} of  interprocedural analysis. However we have  not found any
published  work   which  describes  the  functional   method  for  non
bit-vector problems.

This  work can  be  extended in  many  directions. We  can extend  the
language to  include higher-order functions.  The scope  of the method
can be extended to  include dead-code elimination. If a reference
to the  value of $(\CONS\; e_1\;  e_2)$ is never  used, the expression
need not be evaluated at all.   Our method, in its present form, would
first evaluate  the expression and  then nullify the reference  to it.
The safety of nullification has  to be proven. Finally, the method has
to be implemented to demonstrate its effectiveness.

\vspace{-2mm}

\bibliographystyle{splncs}
\bibliography{fun_hra}
\appendix
\section{Solving Liveness Equations}
\label{sec:solving_eqns}
\newcommand{\nt}{\ensuremath{N}}
\newcommand{\var}[1]{\scalebox{.9}{\mbox{$\langle$#1$\rangle$}}}

In general,  the equations  defining the functions \Lfonly\
will be recursive.  To solve such equations, we start by guessing that
the solution will be of the form:
\begin{eqnarray*}
  \Lf{f}{i}(\sigma) &=& \Uf{f}{i} \plus \Df{f}{i}\cdot\sigma,
\end{eqnarray*}

where \Uf{f}{i}  and \Df{f}{i} are  sets of strings over  the alphabet
$\lbrace  \acar, \acdr,\bcar,  \bcdr \rbrace$.   The  intuition behind
this form  of solution  is as  follows: The function  $f$ can  use its
argument locally  and/or copy a part  of it to the  return value being
computed. \Uf{f}{i} is the set  of live paths of $i^{th}$ argument due
to local use in $f$.  \Df{f}{i} is a sort of selector that selects the
live  paths  corresponding  to  the  $i^{th}$  argument  of  $f$  from
$\sigma$, the liveness paths of the return value.

If  we substitute  the  guessed  form of  \Lf{f}{i}  in the  equations
describing it and  equate the terms containing $\sigma$  and the terms
without   $\sigma$,  we   get  the   equations  for   \Uf{f}{i}  and
\Df{f}{i}. This is illustrated in the following example.

\begin{example}\label{exmp:decompose-eqn}
  Consider   the   equation   for   \Lf{\append}{1}($\sigma$)   from
  Example~\ref{exmp:motivation_analysis}:

$\begin{array}{rcl}
  \Lf{\append}{1}(\sigma)
  &=& \epsilonset \plus \lbrace\acar\bcar\rbrace\cdot\sigma \plus
  \acdrset\cdot\Lf{\append}{1}(\bcdrset\cdot\sigma)
\end{array}$

\noindent  Decomposing both  sides  of the  equation, and  rearranging
gives:

$\begin{array}{rcl}
  \Uf{\append}{1} \plus
  \Df{\append}{1}\cdot\sigma
  &=&
  \epsilonset \plus \acdrset\cdot\Uf{\append}{1} \\
  && \plus\; \lbrace\acar\bcar\rbrace\cdot\sigma \plus
  \acdrset\cdot\Df{\append}{1}\cdot\bcdrset\cdot\sigma
\end{array}$

\noindent  Separating the parts  that are  $\sigma$ dependent  and the
parts that are $\sigma$  independent, and equating them separately, we
get:

$\begin{array}{rcl}
  \Uf{\append}{1} &=& \epsilonset
  \plus \acdrset\cdot\Uf{\append}{1} \\
  \Df{\append}{1}\cdot\sigma &=& \lbrace\acar\bcar\rbrace\cdot\sigma
  \plus \acdrset\cdot\Df{\append}{1}\cdot\bcdrset\sigma \\
  &=& (\lbrace\acar\bcar\rbrace
  \plus \acdrset\cdot\Df{\append}{1}\cdot\bcdrset)\cdot\sigma
\end{array}$

\noindent As the equations hold  for any general $\sigma$, we simplify
them to:

$\begin{array}{rcl}
  \Uf{\append}{1} &=& \epsilonset
  \plus \acdrset\cdot\Uf{\append}{1} \\
  \Df{\append}{1} &=& \lbrace\acar\bcar\rbrace 
  \plus \acdrset\cdot\Df{\append}{1}\cdot\bcdrset
\end{array}$

\noindent     Similarly,      from     the     equation     describing
\Lf{\append}{2}($\sigma$), we get:

$\begin{array}{rcl}
  \Uf{\append}{2} &=& \Uf{\append}{2} \\
  \Df{\append}{2} &=& \epsilonset \plus \Df{\append}{2}\cdot\bcdrset
\end{array}$

\noindent These equations describe the transfer functions for \append.
\hfill\ \qed\end{example}

The values of \Uf{}{} and \Df{}{} 
are sets  of strings over  the alphabet $\lbrace \acar,  \acdr, \bcar,
\bcdr \rbrace$.  
We  are interested  in  least solutions  to  the equations  describing
$\Uf{}{}\!$ and $\Df{}{}\!$. 
We use context  free grammars (CFG) to describe these
solutions.
The set of terminal symbols of the CFG is $\lbrace\acar, \acdr, \bcar,
\bcdr\rbrace$.  Non-terminals and  associated rules are constructed as
illustrated             in             Examples~\ref{exmp:cfg-eqn-app}
and~\ref{exmp:cfg-eqn-full}.

\begin{example}\label{exmp:cfg-eqn-app}
  Consider         the        following         constraint        from
  Example~\ref{exmp:decompose-eqn}:

    $\begin{array}{rcl}
      \Uf{\append}{1} &=& \epsilonset
      \plus \acdrset\cdot\Uf{\append}{1}
    \end{array}$

  \noindent We  add non-terminal  \var{\Uf{\append}{1}}  and the 
    productions with right hand sides directly derived from the constraints:

    $\begin{array}{rcl}
      \var{\Uf{\append}{1}} &\rightarrow& \epsilon \mid
      \acdr\var{\Uf{\append}{1}}
    \end{array}$

  \noindent 
The productions generated from other constraints of
Example~\ref{exmp:decompose-eqn} are:

$\begin{array}{rcl}
  \var{\Df{\append}{1}} &\rightarrow& \acar\bcar \mid
  \acdr\var{\Df{\append}{1}}\bcdr\\
  \var{\Uf{\append}{2}} &\rightarrow& \var{\Uf{\append}{2}} \\
  \var{\Df{\append}{2}} &\rightarrow& \epsilon \mid
   \var{\Df{\append}{2}}\bcdr
\end{array}$

  \noindent 
These productions describe the transfer functions of \append.
\hfill\ \qed
\end{example}

The liveness environment at each program point can be represented as a
CFG with  a start symbol for  every variable.  To do  so, the analysis
starts with \var{$\nt_\exit$},  the non-terminal describing the liveness
 of  the result of the program,  $\sigma_\exit$.
The  productions  for \var{$\nt_\exit$} are:

$\begin{array}{rcl}
  \var{$\nt_\exit$} &\rightarrow& \epsilon \mid \acar \var{$\nt_\exit$}
  \mid\acdr \var{$\nt_\exit$}
\end{array}$

\begin{example}\label{exmp:cfg-eqn-full}
  Let {$\nt_{\pi}^{\sf  v}$} denote the  non-terminal corresponding to
   the liveness  associated with a  variable {\sf v} at  program point
   $\pi$.  For the program of Fig.~\ref{fig:mot-exmp}:

  $\begin{array}[b]{rcl}
    \var{$\nt_{\pib}^{w}$} &\rightarrow&
    \epsilon \mid \acdr \mid \acdr\acar \mid \acdr\acar\acar\var{$\nt_\exit$}\\
    \var{$\nt_{\pia}^{z}$} &\rightarrow&
    \var{\Uf{\append}{2}} \mid
    \var{\Df{\append}{2}} \mid 
    \var{\Df{\append}{2}}\acdr \mid
    \var{\Df{\append}{2}}\acdr\acar\\
    && \mid \var{\Df{\append}{2}}\acdr\acar\acar \var{$\nt_\exit$} \\
    \var{$\nt_{\pia}^{y}$} &\rightarrow&
    \var{\Uf{\append}{1}} \mid 
    \var{\Df{\append}{1}} \mid
    \var{\Df{\append}{1}}\acdr \mid
    \var{\Df{\append}{1}}\acdr\acar \\
    && \mid \var{\Df{\append}{1}}\acdr\acar\acar \var{$\nt_\exit$}
  \end{array}$\hfill  
\qed\end{example}

It is possible that different  paths, which are not in canonical form,
may  reduce to  the  same canonical  path  and hence  encode the  same
information.   We are  interested in  the information  encoded  by the
paths, and therefore  want to check memberships of  canonical paths in
CFGs. However,  the paths described by  the CFGs resulting  out of our
analysis are  not in canonical form.   It is not obvious  how to check
the membership  of canonical  paths directly in  such CFGs.   To solve
this problem, we need equivalent CFGs such that if $\alpha$ belongs to
an original  CFG and  $\alpha \rightstar \beta$,  where $\beta$  is in
canonical form, then $\beta$ belongs to the corresponding new CFG.
Directly converting the reduction rules~(\ref{eqn:eup-red})
into  productions  and  adding  it  to the  grammar  results  in  {\em
unrestricted grammar}~\cite{hopcraft90toc}.   To simplify the problem,
we  approximate  original CFGs  by  non-deterministic finite  automata
(NFAs) and convert them to equivalent NFAs which can be used to check
the membership of canonical paths.

\subsection{Approximating CFGs using NFAs}
\label{sec:nfa-eqn}

The conversion of a CFG $\gram$ to an approximate NFA $\nfa$ should be
safe in that  the language accepted by $\nfa$ should  be a superset of
the language  accepted by $\gram$.  We use the algorithm  described by
Mohri and  Nederhof~\cite{mohri00regular}. The algorithm  transforms a
CFG to a  restricted form called {\em strongly  regular} CFG which can
be converted easily to a finite automaton.

\newcommand{\nfaNN}[1]{%
    \psset{unit=1mm}
  \begin{pspicture}(0,0)(15,13)
    \putnode{n0}{origin}{0}{3}{}
    \putnode{nn}{n0}{10}{0}{
      \pscirclebox[framesep=2,doubleline=true]{\mbox{ }}}
    \ncline[nodesepB=-1]{->}{n0}{nn}\Aput[1]{start}
    \nccurve[angleA=110,angleB=70,ncurv=5,nodesepB=0]{->}{nn}{nn}\Aput[.1]{#1}
  \end{pspicture}
}

\begin{example}\label{exmp:nfa-eqn}
We  show  the  approximate  NFAs  for each  of  the  non-terminals  in
Example~\ref{exmp:cfg-eqn-app} and Example~\ref{exmp:cfg-eqn-full}.

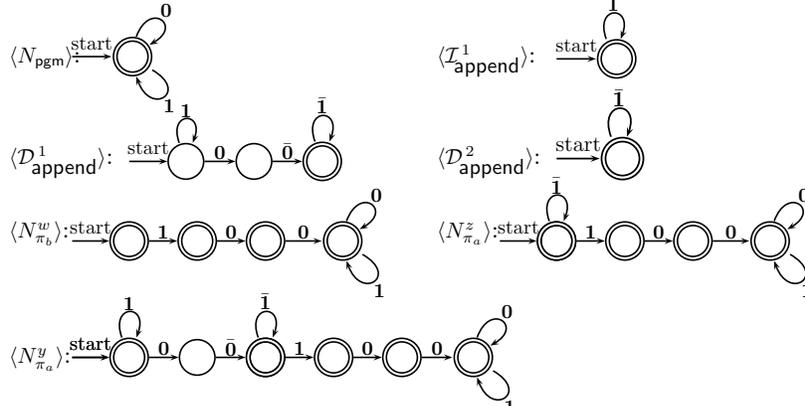
\raisebox{-1cm}{\scalebox{.91}{
\begin{tabular}{@{}l@{\ \ }ll@{\ \ }l@{}}
  \var{$\nt_\exit$}:&    
  \raisebox{-5mm}{\scalebox{.83}{\psset{unit=1mm}
  \begin{pspicture}(0,0)(28,16)
    \putnode{n0}{origin}{-10}{7}{}
    \putnode{na}{n0}{10}{0}{
      \pscirclebox[framesep=2,doubleline=true]{\mbox{ }}}
    \ncline[nodesepB=-1]{->}{n0}{na}\Aput[1]{start}
    \nccurve[angleA=70,angleB=30,ncurv=5,nodesepB=-.5]{->}{na}{na}\Aput[.1]{\acar}
    \nccurve[angleA=-30,angleB=-70,ncurv=5]{->}{na}{na}\Aput[.2]{\acdr}
  \end{pspicture}}}
  &
  \var{\Uf{\append}{1}}:&
  \raisebox{-2mm}{\scalebox{.83}{\nfaNN{\acdr}}}
  \\
  \var{\Df{\append}{1}}:&  
  \raisebox{-3mm}{\scalebox{.83}{\psset{unit=1mm}
  \begin{pspicture}(0,0)(28,15)
    \putnode{n0}{origin}{0}{4}{}
    \putnode{na}{n0}{10}{0}{\pscirclebox[framesep=2.4]{\mbox{ }}}
    \ncline[nodesepB=0]{->}{n0}{na}\Aput[1]{start}
    \nccurve[angleA=110,angleB=70,ncurv=5]{->}{na}{na}\Aput[.1]{\acdr}
    \putnode{nb}{na}{12}{0}{\pscirclebox[framesep=2.4]{\mbox{ }}}
    \putnode{nc}{nb}{12}{0}{\pscirclebox[framesep=2,doubleline=true]{\mbox{ }}}
    \nccurve[angleA=110,angleB=70,ncurv=5]{->}{nc}{nc}\Aput[.2]{\bcdr}

    \ncline[nodesep=0]{->}{na}{nb}\Aput[.1]{\acar}
    \ncline[nodesep=0]{->}{nb}{nc}\Aput[.1]{\bcar}
  \end{pspicture}}}
  &
  \var{\Df{\append}{2}}:& \raisebox{-2mm}{\scalebox{.91}{\nfaNN{\bcdr}}}
  \\

  \raisebox{8mm}{\var{$\nt_{\pib}^{w}$}:} &
{\scalebox{.83}{\psset{unit=1mm}
  \begin{pspicture}(0,0)(53,18)
    \putnode{n0}{origin}{-10}{9}{}
    \putnode{na}{n0}{10}{0}{\pscirclebox[framesep=2,doubleline=true]{\mbox{ }}}
    \ncline[nodesepB=0]{->}{n0}{na}\Aput[1]{start}
    \putnode{nb}{na}{12}{0}{\pscirclebox[framesep=2,doubleline=true]{\mbox{ }}}
    \putnode{nc}{nb}{12}{0}{\pscirclebox[framesep=2,doubleline=true]{\mbox{ }}}

    \ncline[nodesep=0]{->}{na}{nb}\Aput[.1]{\acdr}
    \ncline[nodesep=0]{->}{nb}{nc}\Aput[.1]{\acar}
    \putnode{nd}{nc}{13}{0}{
      \pscirclebox[framesep=2,doubleline=true]{\mbox{ }}}
    \ncline[nodesepB=-1]{->}{nc}{nd}\Aput[.2]{\acar}
    \nccurve[angleA=70,angleB=30,ncurv=5,nodesepB=-.5]{->}{nd}{nd}\Aput[.1]{\acar}
    \nccurve[angleA=-30,angleB=-70,ncurv=5]{->}{nd}{nd}\Aput[.2]{\acdr}
  \end{pspicture}}} 
  &
  \raisebox{8mm}{\var{$\nt_{\pia}^{z}$}:} &
  \multicolumn{1}{@{}l@{}}{\scalebox{.83}{\psset{unit=1mm}
  \begin{pspicture}(0,0)(63,19)
    \putnode{mb}{origin}{-10}{9}{}
    \putnode{ma}{mb}{10}{0}{\pscirclebox[framesep=2,doubleline=true]{\mbox{ }}}
    \ncline[nodesepB=0]{->}{mb}{ma}\Aput[1]{start}
    \nccurve[angleA=110,angleB=70,ncurv=5,nodesepB=0]{->}{ma}{ma}\Aput[.1]{\bcdr}
    \putnode{na}{ma}{0}{0}{\pscirclebox[framesep=2,doubleline=true]{\mbox{ }}}
    \putnode{nb}{na}{12}{0}{\pscirclebox[framesep=2,doubleline=true]{\mbox{ }}}
    \putnode{nc}{nb}{12}{0}{\pscirclebox[framesep=2,doubleline=true]{\mbox{ }}}

    \ncline[nodesep=0]{->}{na}{nb}\Aput[.1]{\acdr}
    \ncline[nodesep=0]{->}{nb}{nc}\Aput[.1]{\acar}
    \putnode{nd}{nc}{13}{0}{
      \pscirclebox[framesep=2,doubleline=true]{\mbox{ }}}
    \ncline[nodesepB=-1]{->}{nc}{nd}\Aput[.2]{\acar}
    \nccurve[angleA=70,angleB=30,ncurv=5,nodesepB=-.5]{->}{nd}{nd}\Aput[.1]{\acar}
    \nccurve[angleA=-30,angleB=-70,ncurv=5]{->}{nd}{nd}\Aput[.2]{\acdr}
  \end{pspicture}}}
  \\
%
  \raisebox{5mm}{\var{$\nt_{\pia}^{y}$}:} &
  \multicolumn{3}{@{}l@{}}{\scalebox{.83}{\psset{unit=1mm}
  \begin{pspicture}(0,0)(45,17)
    \putnode{n0}{origin}{-10}{7}{}
    \putnode{na}{n0}{10}{0}{\pscirclebox[framesep=2,doubleline=true]{\mbox{ }}}
    \ncline[nodesepB=0]{->}{n0}{na}\Aput[1]{start}
    \nccurve[angleA=110,angleB=70,ncurv=5]{->}{na}{na}\Aput[.1]{\acdr}
    \putnode{nb}{na}{12}{0}{\pscirclebox[framesep=2.4]{\mbox{ }}}
    \putnode{nc}{nb}{12}{0}{\pscirclebox[framesep=2,doubleline=true]{\mbox{ }}}
    \nccurve[angleA=110,angleB=70,ncurv=5]{->}{nc}{nc}\Aput[.2]{\bcdr}
    \ncline[nodesep=0]{->}{na}{nb}\Aput[.1]{\acar}
    \ncline[nodesep=0]{->}{nb}{nc}\Aput[.1]{\bcar}
    
    \putnode{ma}{nc}{0}{0}{\pscirclebox[framesep=2,doubleline=true]{\mbox{ }}}
    \ncline[nodesepB=0]{->}{n0}{na}\Aput[1]{start}
    \putnode{mb}{ma}{12}{0}{\pscirclebox[framesep=2,doubleline=true]{\mbox{ }}}
    \putnode{mc}{mb}{12}{0}{\pscirclebox[framesep=2,doubleline=true]{\mbox{ }}}

    \ncline[nodesep=0]{->}{ma}{mb}\Aput[.1]{\acdr}
    \ncline[nodesep=0]{->}{mb}{mc}\Aput[.1]{\acar}
    \putnode{md}{mc}{12}{0}{
      \pscirclebox[framesep=2,doubleline=true]{\mbox{ }}}
    \ncline[nodesepB=-1]{->}{mc}{md}\Aput[.2]{\acar}
    \nccurve[angleA=70,angleB=30,ncurv=5,nodesepB=-.5]{->}{md}{md}\Aput[.1]{\acar}
    \nccurve[angleA=-30,angleB=-70,ncurv=5]{->}{md}{md}\Aput[.2]{\acdr}
  \end{pspicture}}}
\end{tabular}
}}
\noindent     Note    that     there    is     no     automaton    for
\var{\Uf{\append}{2}}. This  is because the least  solution of the
equation          \mbox{$\var{\Uf{\append}{2}}         \rightarrow
\var{\Uf{\append}{2}}$}   is  $\emptyset$.   Also,   the  language
accepted by the automaton for \Df{\append}{1} is approximate as it
does not ensure that there is  an equal number of \acdr\ and \bcdr\ in
the strings generated by rules for \var{\Df{\append}{1}}.
\hfill\ \qed\end{example} 

\subsection{Conversion of NFAs to Accept Canonical Paths}
\label{sec:nfa-elim}

{ Algorithm~\ref{algo:simplify-nfa}  converts an NFA  with transitions
  on  symbols \bcar\  and  \bcdr\  to an  equivalent  NFA without  any
  transitions  on these symbols.  The algorithm  repeatedly introduces
  $\epsilon$  edges to  bypass  a pair  of  consecutive edges  labeled
  \bcar\acar\ or  \bcdr\acdr.  The process  is continued till  a fixed
  point is  reached.  When the  fixed point is reached,  the resulting
  NFA contains the  canonical paths corresponding to all  the paths in
  the original  NFA.  The paths not  in canonical form  are deleted by
  removing edges labeled \bcar\ and \bcdr.  Note that by our reduction
  rules if  $\alpha$ is accepted  by $\nfabar$ and  $\alpha \rightstar
  \bot$,  then $\bot$ should  be accepted  by $\nfa$,  However, $\nfa$
  returned by  our algorithm does not  accept $\bot$.}  This  is not a
problem  because the  paths which  are tested  for  membership against
$\nfa$ do not include $\bot$ as well.
\begin{algorithm}[t]
\caption{Simplifying NFA}\label{algo:simplify-nfa}
\noindent  {\bf  Input:}  An NFA  \nfabar\  with  underlying  alphabet
  $\lbrace  \acar,  \acdr, \bcar,  \bcdr\rbrace$  \\
  {\bf Output:} An NFA \nfa\ with underlying  alphabet $\lbrace \acar,
  \acdr\rbrace$  accepting  the  equivalent  set of  paths
  \\
  {\bf Steps:}
\begin{algorithmic}
\footnotesize
  \STATE $i \leftarrow 0$ 
  \STATE $\nfa_0 \leftarrow$ Equivalent NFA of \nfabar\  without
  $\epsilon$-moves \cite{hopcraft90toc}
  
  \REPEAT
  \STATE $\nfa'_{i+1} \leftarrow \nfa_i$
  \FORALL{states $q$ in $\nfa_i$ such that $q$ has an incoming
    edge from $q'$ with label $\bcar$ and outgoing edge to $q''$ with label
    $\acar$} 
  \STATE add an edge in $\nfa'_{i+1}$ from $q'$ to $q''$ with label
  $\epsilon$.
  \COMMENT{bypass  $\bcar\acar$ using $\epsilon$}
  \ENDFOR
  
  \FORALL{states $q$ in $\nfa_i$ such that $q$ has an incoming
    edge from $q'$
    with label $\bcdr$ and outgoing edge to $q''$ with label
    $\acdr$}
  \STATE add an edge in $\nfa'_{i+1}$ from $q'$ to $q''$ with label
  $\epsilon$.
  \COMMENT{bypass  $\bcdr\acdr$ using $\epsilon$}
  \ENDFOR
  
  \STATE $\nfa_{i+1} \leftarrow$ Equivalent NFA of $\nfa'_{i+1}$ without
  $\epsilon$-moves
  
  \STATE $i \leftarrow i+1$
  \UNTIL ($\nfa_{i} = \nfa_{i-1}$)
  
  \STATE $\nfa \leftarrow \nfa_i$
  \STATE delete all edges with label \bcar\ or \bcdr\ in \nfa.
\end{algorithmic}
\end{algorithm}

\begin{example}\label{exmp:elim-01}
We  show the elimination  of \bcar\  and \bcdr\  for the  automata for
\var{$\nt_{\pia}^{y}$}    and   \var{$\nt_{\pia}^{z}$}.    The
automaton for \var{$\nt_{\pib}^{w}$}  remains unchanged as it does
not  contain transitions  on \bcar\  and \bcdr.   The automata  at the
termination of the loop in the algorithm are:

\hspace{-8mm}
\scalebox{.91}{
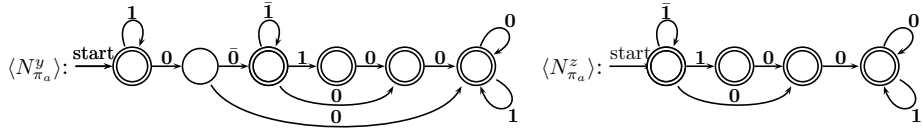
  \begin{tabular}{@{}ll@{}ll@{}}
    \raisebox{08mm}{\var{$\nt_{\pia}^{y}$}:} &
\scalebox{.83}{\psset{unit=1mm}
  \begin{pspicture}(0,0)(82,23)
    \putnode{n0}{origin}{0}{11}{}
    \putnode{na}{n0}{10}{0}{\pscirclebox[framesep=2,doubleline=true]{\mbox{ }}}
    \ncline[nodesepB=0]{->}{n0}{na}\Aput[1]{start}
    \nccurve[angleA=110,angleB=70,ncurv=5]{->}{na}{na}\Aput[.1]{\acdr}
    \putnode{nb}{na}{12}{0}{\pscirclebox[framesep=2.4]{\mbox{ }}}
    \putnode{nc}{nb}{12}{0}{\pscirclebox[framesep=2,doubleline=true]{\mbox{ }}}
    \nccurve[angleA=110,angleB=70,ncurv=5]{->}{nc}{nc}\Aput[.2]{\bcdr}
    \ncline[nodesep=0]{->}{na}{nb}\Aput[.1]{\acar}
    \ncline[nodesep=0]{->}{nb}{nc}\Aput[.1]{\bcar}
    
    \putnode{ma}{nc}{0}{0}{\pscirclebox[framesep=2,doubleline=true]{\mbox{ }}}
    \ncline[nodesepB=0]{->}{n0}{na}\Aput[1]{start}
    \putnode{mb}{ma}{12}{0}{\pscirclebox[framesep=2,doubleline=true]{\mbox{ }}}
    \putnode{mc}{mb}{12}{0}{\pscirclebox[framesep=2,doubleline=true]{\mbox{ }}}

    \ncline[nodesep=0]{->}{ma}{mb}\Aput[.1]{\acdr}
    \ncline[nodesep=0]{->}{mb}{mc}\Aput[.1]{\acar}
    \putnode{md}{mc}{12}{0}{
      \pscirclebox[framesep=2,doubleline=true]{\mbox{ }}}
    \ncline[nodesepB=-1]{->}{mc}{md}\Aput[.2]{\acar}
    \nccurve[angleA=70,angleB=30,ncurv=5,nodesepB=-.5]{->}{md}{md}\Aput[.1]{\acar}
    \nccurve[angleA=-30,angleB=-70,ncurv=5]{->}{md}{md}\Aput[.2]{\acdr}
    \nccurve[angleA=-60,angleB=-120,ncurv=.5]{->}{nc}{mc}\Aput[.2]{\acar}
    \nccurve[angleA=-60,angleB=-120,ncurv=.5]{->}{nb}{md}\Aput[.2]{\acar}
  \end{pspicture}} &
  \raisebox{08mm}{\var{$\nt_{\pia}^{z}$}:} &
\scalebox{.83}{\psset{unit=1mm}
  \begin{pspicture}(0,0)(63,23)
    \putnode{mb}{origin}{0}{11}{}
    \putnode{ma}{mb}{10}{0}{\pscirclebox[framesep=2,doubleline=true]{\mbox{ }}}
    \ncline[nodesepB=-1]{->}{mb}{ma}\Aput[1]{start}
    \nccurve[angleA=110,angleB=70,ncurv=5,nodesepB=0]{->}{ma}{ma}\Aput[.1]{\bcdr}
    \putnode{na}{ma}{0}{0}{\pscirclebox[framesep=2,doubleline=true]{\mbox{ }}}
    \putnode{nb}{na}{12}{0}{\pscirclebox[framesep=2,doubleline=true]{\mbox{ }}}
    \putnode{nc}{nb}{12}{0}{\pscirclebox[framesep=2,doubleline=true]{\mbox{ }}}

    \ncline[nodesep=0]{->}{na}{nb}\Aput[.1]{\acdr}
    \ncline[nodesep=0]{->}{nb}{nc}\Aput[.1]{\acar}
    \putnode{nd}{nc}{13}{0}{
      \pscirclebox[framesep=2,doubleline=true]{\mbox{ }}}
    \ncline[nodesepB=-1]{->}{nc}{nd}\Aput[.2]{\acar}
    \nccurve[angleA=70,angleB=30,ncurv=5,nodesepB=-.5]{->}{nd}{nd}\Aput[.1]{\acar}
    \nccurve[angleA=-30,angleB=-70,ncurv=5]{->}{nd}{nd}\Aput[.2]{\acdr}
    \nccurve[angleA=-60,angleB=-120,ncurv=.5]{->}{ma}{nc}\Aput[.2]{\acar}
  \end{pspicture}}
\end{tabular}
}

\noindent Eliminating the edges labeled \bcar\ and \bcdr, and removing
the dead states gives:

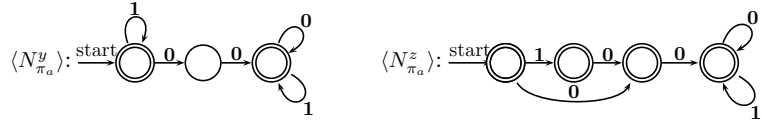
\hfill\scalebox{.91}{
  \begin{tabular}{ll@{\hspace{10mm}}ll}
    \raisebox{6mm}{\var{$\nt_{\pia}^{y}$}:} 
    &
    \scalebox{.83}{\psset{unit=1mm}
      \begin{pspicture}(0,0)(40,19)
	\putnode{n0}{origin}{0}{8}{}
	\putnode{na}{n0}{10}{0}{\pscirclebox[framesep=2,doubleline=true]{\mbox{ }}}
	\ncline[nodesepB=0]{->}{n0}{na}\Aput[1]{start}
	\nccurve[angleA=110,angleB=70,ncurv=5]{->}{na}{na}\Aput[.1]{\acdr}
	\putnode{nb}{na}{12}{0}{\pscirclebox[framesep=2.4]{\mbox{ }}}
	\ncline[nodesep=0]{->}{na}{nb}\Aput[.1]{\acar}
	\putnode{md}{nb}{12}{0}{\pscirclebox[framesep=2,doubleline=true]{\mbox{ }}}
	\nccurve[angleA=70,angleB=30,ncurv=5,nodesepB=-.5]{->}{md}{md}\Aput[.1]{\acar}
	\nccurve[angleA=-30,angleB=-70,ncurv=5]{->}{md}{md}\Aput[.2]{\acdr}
	\ncline{->}{nb}{md}\Aput[.2]{\acar}
      \end{pspicture}
    }
    &
    \raisebox{6mm}{\var{$\nt_{\pia}^{z}$}:} 
    &
    \scalebox{.83}{\psset{unit=1mm}
      \begin{pspicture}(0,0)(63,19)
	\putnode{mb}{origin}{0}{8}{}
	\putnode{ma}{mb}{10}{0}{\pscirclebox[framesep=2,doubleline=true]{\mbox{ }}}
	\ncline[nodesepB=-1]{->}{mb}{ma}\Aput[1]{start}
	\putnode{na}{ma}{0}{0}{\pscirclebox[framesep=2,doubleline=true]{\mbox{ }}}
	\putnode{nb}{na}{12}{0}{\pscirclebox[framesep=2,doubleline=true]{\mbox{ }}}
	\putnode{nc}{nb}{12}{0}{\pscirclebox[framesep=2,doubleline=true]{\mbox{ }}}
	
	\ncline[nodesep=0]{->}{na}{nb}\Aput[.1]{\acdr}
	\ncline[nodesep=0]{->}{nb}{nc}\Aput[.1]{\acar}
	\putnode{nd}{nc}{13}{0}{
	  \pscirclebox[framesep=2,doubleline=true]{\mbox{ }}}
	\ncline[nodesepB=-1]{->}{nc}{nd}\Aput[.2]{\acar}
	\nccurve[angleA=70,angleB=30,ncurv=5,nodesepB=-.5]{->}{nd}{nd}\Aput[.1]{\acar}
	\nccurve[angleA=-30,angleB=-70,ncurv=5]{->}{nd}{nd}\Aput[.2]{\acdr}
	\nccurve[angleA=-60,angleB=-120,ncurv=.5]{->}{ma}{nc}\Aput[.2]{\acar}
    \end{pspicture}}
  \end{tabular}
}\hfill\mbox{}

The  language accepted  by these  automata represent  the  live access
paths corresponding to $y$ and $z$ at $\pia$.
\qed\end{example}

We now give the proofs of the termination and correctness of our algorithm.

\newcommand{\mylongrightarrow}{
  \ \pnode{p0}\hspace{10.5mm}\pnode{p1}\ %
  \ncline{->}{p0}{p1}
}

\newcommand{\rightarroweps}{\mbox{\raisebox{1.2mm}{\mylongrightarrow
   \Aput[.1]{\footnotesize addition}
   \Bput[.1]{\footnotesize of $\epsilon$-edges}
}}}
\newcommand{\rightarrownoeps}{\mbox{\raisebox{1.2mm}{\mylongrightarrow
   \Aput[.1]{\footnotesize deletion}
   \Bput[.1]{\footnotesize of $\epsilon$-edges}
}}}

\newcommand{\rightarrownobar}{\mbox{\raisebox{1.2mm}{\mylongrightarrow
   \Aput[.1]{\footnotesize deletion of }
   \Bput[.1]{\footnotesize \bcar, \bcdr\ edges}
}}}

\subsubsection*{Termination}
Termination  of  the  algorithm  follows  from  the  fact  that  every
iteration of {\bf do-while} loop adds  new edges to the NFA, while old
edges are not  deleted.  Since no new states are added  to NFA, only a
fixed number of  edges can be added before we reach  a fix point.

\subsubsection*{Correctness}
The sequence of obtaining \nfa\ from \nfabar\
can be viewed as follows, with $\nfa_m$ denoting the NFA at the
termination of while loop:

$$    \nfabar   \rightarrownoeps    \nfa_0    \rightarroweps
\cdots 
\nfa'_i \rightarrownoeps  \nfa_i
\cdots  \rightarrownoeps  \nfa_m$$ 

$$\nfa_m \rightarrownobar \nfa$$\\[-2mm]

\noindent  Then,  the  languages  accepted  by  these  NFAs  have  the
following relation:

$$ L(\nfabar)  = L(\nfa_0) \subseteq
\cdots \subseteq L(\nfa'_i) = L(\nfa_i) \subseteq \cdots = L(\nfa_m)$$
$$ L(\nfa) \subseteq L(\nfa_m) $$

We first prove that the addition of $\epsilon$-edges in the while loop
does not add any new information, i.e.\ any path accepted by
the NFA after the addition of $\epsilon$-edges is a reduced version of
some  path existing  in  the  NFA  before the  addition  of
$\epsilon$-edges.

\begin{lemma}\label{lemma:no-new}
  for $i > 0$, if $\alpha \in L(\nfa_i)$ then there exists $\alpha'
  \in L(\nfa_{i-1})$ such that $\alpha' \rightstar \alpha$.
\end{lemma}
\begin{proof}
  As $ L(\nfa_i) = L(\nfa'_i)$, we have $\alpha \in L(\nfa'_i)$. Only
  difference between $\nfa'_i$ and $\nfa_{i-1}$ is that $\nfa'_i$
  contains some extra $\epsilon$-edges. Thus, any $\epsilon$-edge free
  path in $\nfa'_i$ is also in $\nfa_{i-1}$. Consider a path $p$ in
  $\nfa'_i$ that accepts $\alpha$. Assume the number of $\epsilon$
  edges in $p$ is $k$. The proof is by induction on $k$.\\
  \noindent{({\em BASE})} $k = 0$, i.e.\ $p$ does not contains any
    $\epsilon$-edge: As the path $p$ is $\epsilon$-edge free, it must
    be present in $\nfa_{i-1}$. Thus, $\nfa_{i-1}$ also accepts
    $\alpha$. $\alpha \rightstar \alpha$.\\
  \noindent{({\em HYPOTHESIS})} For any $\alpha \in L(\nfa_i) $ with
    accepting path $p$ having less than $k$ $\epsilon$-edges there
    exists $\alpha' \in L(\nfa_{i-1})$ such that $\alpha' \rightstar
    \alpha$. \\
    \noindent{({\em  INDUCTION})}  $p$  contains $k$  $\epsilon$-edges
    $e_1,\ldots,e_k$: Assume  $e_1$ connects states $q'$  and $q''$ in
    $\nfa'_i$. By construction, there  exists a state $q$ in $\nfa'_i$
    such that  there is  an edge  $e'_1$ from $q'$  to $q$  with label
    \bcar(\bcdr)  and an  edge $e''_1$  from $q$  to $q''$  with label
    \acar(\acdr) in $\nfa'_i$. Replace  $e_1$ by $e'_1e''_1$ in $p$ to
    get a  new path  $p''$ in $\nfa'_i$.   Let $\alpha''$ be  the path
    accepted by  $p''$. Clearly, $\alpha''  \rightk{1} \alpha$.  Since
    $p''$  has  $k-1$  $\epsilon$-edges,  $\alpha''$  is  accepted  by
    $\nfa'_i$   along  a  path   ($p''$)  that   has  less   than  $k$
    $\epsilon$-edges.  By  induction hypothesis, we  have $\alpha' \in
    L(\nfa_{i-1})$ such  that $\alpha'\rightstar\alpha''$.  This along
    with $\alpha'' \rightk{1} \alpha$ gives $\alpha'\rightstar\alpha$.
\end{proof}

\begin{corollary}\label{corr:no-new}
  for each $\alpha \in L(\nfa_m)$, there exists $\alpha' \in
  L(\nfabar)$ such that $\alpha' \rightstar \alpha$.
\end{corollary}
\begin{proof}
  The    proof    is    by     induction    on    $m$,    and    using
  Lemma~\ref{lemma:no-new}.
\end{proof} 

The following lemma  shows that the the language  accepted by $\nfa_m$
is closed with respect to reduction of paths.

\begin{lemma}\label{lemma:closure}
  For $\alpha \in L(\nfa_m)$, if $\alpha \rightstar
  \alpha'$ and $\alpha' \not= \bot$, then $\alpha' \in  L(\nfa_m)$.
\end{lemma}
\begin{proof}
   Assume $\alpha \rightk{k} \alpha'$. The Proof is by induction on
  $k$, number of steps in reduction.\\
    \noindent{({\em   BASE})}  case  $k=0$   is  trivial   as  $\alpha
      \rightk{0} \alpha$.\\
    \noindent{({\em   HYPOTHESIS})}  Assume   that  for   $\alpha  \in
      L(\nfa_m)$, if $\alpha  \rightk{k-1} \alpha'$, then $\alpha' \in
      L(\nfa_m)$.\\
    \noindent{({\em  INDUCTION})}   $\alpha  \in  L(\nfa_m)$,  $\alpha
      \rightk{k} \alpha'$. There  exists $\alpha''$ such that: $\alpha
      \rightk{k-1}\;  \alpha''\;  \rightk{1}  \alpha'$.  By  induction
      hypothesis, we have $\alpha'' \in L(\nfa_m)$.

      For $\alpha'' \rightk{1} \alpha'$ to hold we must have $\alpha''
      = \alpha_1\bcar\acar\alpha_2$  and $\alpha' = \alpha_1\alpha_2$,
      or  $\alpha''  =   \alpha_1\bcdr\acdr\alpha_2$  and  $\alpha'  =
      \alpha_1\alpha_2$.   Consider   the   case  when   $\alpha''   =
      \alpha_1\bcar\acar\alpha_2$.   Any  path  in $\nfa_m$  accepting
      $\alpha''$ must have the  following structure ({The states shown
      separately may not necessarily be different}):
      \begin{center}
	\scalebox{.91}
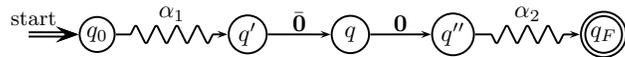
{\begin{pspicture}(0,0)(10,.8)
	\psset{unit=1mm}
	\putnode{s0}{origin}{3}{4}{}
	\putnode{s1}{s0}{10}{0}{\pscirclebox[framesep=.5]{$q_0$}}
	\putnode{s2}{s1}{22}{0}{\pscirclebox[framesep=.5]{$q'$}}
	\putnode{s3}{s2}{15}{0}{\pscirclebox[framesep=1.2]{$q$}}
	\putnode{s4}{s3}{15}{0}{\pscirclebox[framesep=.5]{$q''$}}
	\putnode{s5}{s4}{22}{0}{\pscirclebox[doubleline=true,
	    framesep=.3]{$q_F$}}
	\ncline[doubleline=true]{->}{s0}{s1}
	\aput[.5](.1){\footnotesize start}
	\ncline{->}{s2}{s3}
	\Aput[.2]{\bcar}
	\ncline{->}{s3}{s4}
	\Aput[.2]{\acar}
	\nczigzag[coilarm=2,coilwidth=3,linearc=.2]{->}{s1}{s2}
	\Aput[1.7]{$\alpha_1$}
	\nczigzag[coilarm=2,coilwidth=3,linearc=.2]{->}{s4}{s5}
	\Aput[1.7]{$\alpha_2$}
      \end{pspicture}}
      \end{center}
      As $\nfa_m$  is the  fixed point NFA  for the  iteration process
      described    in    the algorithm,   adding    an
      $\epsilon$-edge between  states $q'$  and $q''$ will  not change
      the  language  accepted by  $\nfa_m$.  But,  the path
      accepted after adding  an $\epsilon$-edge is $\alpha_1\alpha_2 =
      \alpha'$. Thus, $\alpha' \in L(\nfa_m)$. The case when $\alpha''
      = \alpha_1\bcdr\acdr\alpha_2$ is identical.
\end{proof}

\begin{corollary}\label{corr:closure}
  For  $\alpha \in  L(\nfabar)$,  if $\alpha  \rightstar \alpha'$  and
  $\alpha' \not= \bot$, then $\alpha' \in L(\nfa_m)$.
\end{corollary}
\begin{proof}
  $L(\nfabar)    \subseteq    L(\nfa_m)    \Rightarrow   \alpha    \in
  L(\nfa_m)$.     The     proof      follows     from
  Lemma~\ref{lemma:closure}.
\end{proof}

The following  theorem asserts the  equivalence of \nfabar\  and \nfa\
with respect to the equivalence of paths, i.e.\ every path in \nfabar\
has an equivalent canonical path in \nfa, and for every canonical path
in \nfa, there exists an equivalent path in \nfabar.

\begin{theorem}\label{thm:equiv-nfa}
  Let  \nfabar\ be  an NFA with  underlying  alphabet $\lbrace  \acar,
  \acdr,  \bcar,  \bcdr\rbrace$.   Let  NFA  \nfa\  be  the  NFA  with
  underlying  alphabet $\lbrace \acar,  \acdr\rbrace$ returned  by the
  algorithm.  Then,
  \begin{enumerate}
  \item if $\alpha  \in L(\nfabar)$, $\beta$ is a  canonical path such
    that $\alpha\rightstar\beta$  and $\beta \not=  \bot$, then $\beta
    \in L(\nfa)$.
  \item if  $\beta \in L(\nfa)$ then  there exists a  path $\alpha \in
    L(\nfabar)$ such that $\alpha\rightstar\beta$.
  \end{enumerate}
\end{theorem}

\begin{proof}
\mbox{}\\
  \begin{enumerate}
  \item  From  Corollary~\ref{corr:closure}:\\  \mbox{$\alpha \in  L(\nfabar),
    \alpha\rightstar\beta$}  and \mbox{$\beta \not= \bot \Rightarrow
    \beta \in L(\nfa_m)$}. As $\beta$ 
    is  in canonical  form,  the path  accepting  $\beta$ in  $\nfa_m$
    consists of edges labeled $\acar$  and $\acdr$ only. The same path
    exists in  \nfa. Thus \nfa\ also accepts  $\beta \Rightarrow \beta
    \in L(\nfa)$.
  \item   $L(\nfa)   \subseteq   L(\nfa_m)   \Rightarrow   \beta   \in
    L(\nfa_m)$.   Using   Corollary~\ref{corr:no-new},  there   exists
    $\alpha \in L(\nfabar)$ such that $\alpha\rightstar\beta$.
  \end{enumerate}
\end{proof}
\end{document}